\documentclass[aps,prd,onecolumn,superscriptaddress,nofootinbib,showpacs]{revtex4}

\usepackage{graphicx}
\usepackage{slashed}
\usepackage{amsmath,bbm,latexsym,amssymb}
\usepackage{epsfig}
\usepackage{epstopdf}

\newcommand{\be}{\begin{equation}}
\newcommand{\ee}{\end{equation}}
\newcommand{\ba}{\begin{eqnarray}}
\newcommand{\ea}{\end{eqnarray}}
\newcommand{\bs}{\begin{subequations}}
\newcommand{\es}{\end{subequations}}

\newcommand{\ra}{\rightarrow}
\newcommand{\Sh}{S}

\newcommand{\grts}{\raise.3ex\hbox{$>$\kern-.75em\lower1ex\hbox{$\sim$}}}
\newcommand{\lets}{\raise.3ex\hbox{$<$\kern-.75em\lower1ex\hbox{$\sim$}}}

\begin{document}
\vspace*{1cm}


\title{Self-cancelation of a scalar in neutral meson mixing and
implications for LHC}

\author{M. Nebot}\thanks{E-mail: nebot@cftp.ist.utl.pt}
\affiliation{CFTP, Departamento de F\'{\i}sica,
Instituto Superior T\'{e}cnico, Universidade de Lisboa,
Avenida Rovisco Pais 1, 1049 Lisboa, Portugal}
\author{Jo\~{a}o P.\ Silva}\thanks{E-mail: jpsilva@cftp.ist.utl.pt}
\affiliation{CFTP, Departamento de F\'{\i}sica,
Instituto Superior T\'{e}cnico, Universidade de Lisboa,
Avenida Rovisco Pais 1, 1049 Lisboa, Portugal}

\date{\today}

\begin{abstract}
Flavour changing neutral scalar interactions are a standard
feature of generic multi Higgs models.
These are constrained by mixing in the neutral meson systems.
We consider situations where there are natural cancelations
in such contributions.
In particular, when the spin 0
particle has both scalar and pseudoscalar couplings,
one may have a \textit{self-cancelation}.
We illustrate one such partial cancelation with
BGL models.
We also inquire whether the flavour changing quark
interactions can lead to new production mechanisms
for a neutral scalar at LHC.
\end{abstract}


\maketitle

\section{Introduction}

The recent discovery of a scalar particle by
ATLAS \cite{ATLASHiggs} and CMS \cite{CMSHiggs} leaves a fundamental question unanswered:
given that a fundamental scalar exists, how many fundamental
scalars are there in Nature?
Indeed,
within an $SU(2)_L \times U(1)_Y$
gauge theory,
the number of gauge bosons is fixed.
In addition,
measurements of the invisible width of the
$Z$ boson at LEP \cite{ALEPH:2005ab} have fixed the number
of fermion families to be three (below $M_Z/2$).
So, only the number of spin zero particles remains to be determined.

The most natural simple extension of the standard model (SM) is the
two Higgs doublet model (2HDM) --
for a review, see for example \cite{hhg, ourreview}.
This model is interesting in itself, but also
as a toy model for a wide number of features that
could appear in other more complicated settings:
among others,
it includes the need to restrict the parameter space,
such that the vacuum does not violate charge;
the appearance of extra scalar and pseudoscalar
particles;
the possibility that CP is violated in the scalar sector,
either explicitly or spontaneously;
the existence of charged scalars which, like
the $W^\pm$, would change flavour;
and the possibility that there are
flavour changing neutral scalar interactions (FCNSI).
In this article we explore the last feature,
considering two consequences.
If there exists a scalar with FCNSI,
this will have a twofold effect.
On the one hand, it could conceivably be produced at
the LHC through quark level interactions such
as $u \bar{c}$, $d \bar{s}$, $\bar{d} s$,
etc\dots
On the other hand,
these production mechanisms must be constrained by the
fact that the same couplings could originate
FCNSI in neutral meson systems,
such as $K^0 - \overline{K}^0$ or $B^0 - \overline{B}^0$,
which we denote generically by $P^0 - \overline{P}^0$.

Flavour changing neutral interactions, such
as $K^0 - \overline{K}^0$ and
$K^0 \ra \mu^+ \mu^-$ have played a crucial role in the
history of Physics.
For example,
CP violation was first seen in the  $K^0 - \overline{K}^0$
system \cite{Christenson:1964fg},
while the charm quark was invented to curtail a
large contribution for $K^0 \ra \mu^+ \mu^-$ from the
box diagram with the up quark \cite{Glashow:1970gm}.
More recently,
the first evidence of CP violation outside of the
kaon system was found by the Babar and Belle experiments
\cite{Aubert:2001nu, Abe:2001xe},
showing that it arises from a large (not small)
parameter,
and the first evidence of $B_{d,\,  s} \rightarrow \mu^+ \mu^-$
has been found \cite{CMS:2014xfa}.

Bounds on $P^0-\overline{P^0}$ mixing
lead to constraints on the coupling $c_{\Sh q_2 q_1}$ of the scalar
$\Sh$ with quarks $q_1$ and $q_2$ that scale linearly
with the mass $m_\Sh $ of $\Sh $:
\be
c_{\Sh  q_2 q_1} \propto  m_\Sh .
\label{lin}
\ee
%
Limits also arise from bounds such as $B_s \rightarrow \mu^+ \mu^-$,
scaling quadratically as
\be
c_{\Sh  q_2 q_1} \propto  \frac{1}{c_{\Sh  \mu \mu}}\ m_\Sh ^2.
\label{quad}
\ee
Generically,
Eq.~\eqref{quad} provides a looser bound than Eq.~\eqref{lin} for
a sufficiently large $m_\Sh $.
Moreover, Eq.~\eqref{quad} involves the coupling $c_{\Sh  \mu \mu}$
of $\Sh $ with $\mu^+ \mu^-$,
which might even be zero if the scalar $\Sh $ has no overlap with
the (would-be) SM Higgs field $H^\textrm{sm}$.
In both cases,
scalars with larger masses have a looser
constraint on $c_{\Sh  q_2 q_1}$ and, thus,
could conceivably be produced at LHC via
$pp \ra q_2 q_1 \ra \Sh $.
This analysis points to a (further) interesting complementary feature
between Flavour Physics, such as would be pursued at
a Super B-factory, and the scalar search to be continued
at LHC's Run2.
Of fundamental interest is the question:
how would $pp \ra q_2 q_1 \ra \Sh $ compare with the glue-glue
production mechanism $pp \ra gg \ra \Sh $?

In Section~\ref{sec:self},
we introduce the mechanism of self-cancelation,
possible when there are both scalar and pseudoscalar couplings.
This is then discussed in increasingly particular
cases of the two most general 2HDM,
the CP conserving 2HDM,
and the BGL models.
In Section~\ref{sec:prod} we turn to the possibility
that the scalar is produced at the LHC through
FCNSI couplings,
and we conclude in section~\ref{sec:conclusions}.
For completeness,
the appendix contains formulas for FCNSI effects
obtained for the most general Lagrangian,
other than those derived in the main text.

\section{\label{sec:self}Self-cancelation in neutral meson mixing}

\subsection{\label{subsec:formulae}Generic scalar contribution}

Let us consider a spin 0 particle $\Sh $ interacting with two quarks
$q_i$ and $q_j$ according to
\be
-{\cal L}_Y
=
\Sh  \left\{
\bar{q}_j\, (a_{{\bar{j}}i} + i b_{{\bar{j}}i} \gamma_5)\, q_i
+
\bar{q}_i\, (a_{{\bar{j}}i}^\ast + i b_{{\bar{j}}i}^\ast \gamma_5)\, q_j
\right\} + \cdots
\label{L}
\ee
No sum on $i$ and $j$ is implied.
If $q_j=q_i$,
then $a$ and $b$ are real;
otherwise, they are complex.
In writing Eq.~\eqref{L} we have already used hermiticity
in the form
\be
a_{{\bar{i}}j} = a_{{\bar{j}}i}^\ast,
\ \ \ \ \
b_{{\bar{i}}j} = b_{{\bar{j}}i}^\ast.
\label{hermiticity}
\ee
If all $b_{ij}$ ($a_{ij}$) were zero,
then $\Sh $ would be a pure scalar (pseudoscalar).
Otherwise,
$\Sh $ will be a mixture of scalar and pseudoscalar,
and there is P violation.
It is interesting to note that there are still viable models
in which the 125 GeV scalar found at LHC has a pure scalar couplings to
the up quarks, while it has a pure pseudoscalar couplings to the down
quarks \cite{Fontes:2015mea}.

We are interested in the neutral meson systems constituted
by $P^0 \sim \bar{q}_j q_i$  and $\overline{P^0} \sim \bar{q}_i q_j$.
${\cal L}_Y$ contributes to an effective Hamiltonian,
mediating the mixing \cite{BLS}
\ba
M_{21}
&=&
\langle \overline{P^0} \left| {\cal H}_\textrm{eff} \right| P^0 \rangle.
\nonumber\\
&=&
\frac{f_P^2\, m_P}{24\, m_\Sh ^2}
\left[
a_{{\bar{j}}i}^2 K_a^P + b_{{\bar{j}}i}^2 K_b^P
\right].
\label{M21}
\ea
We denote by $m_P$, $\Delta m_P$, and $f_P$ the average mass,
the mass difference,
and the form factor of the $P^0-\overline{P^0}$ system,
respectively.
Under reasonable approximations \cite{BLS},
\be
\Delta m_P = 2 |M_{21}|.
\label{DeltaM}
\ee
In the vacuum insertion approximation,
discussed in detail in appendix C
of Ref.~\cite{BLS}\footnote{Notice that there are
a few sign misprints in the hardcover edition of \cite{BLS},
corrected both in the paperback edition and here.},
we find
\ba
K_a^P &=& 1 + \frac{m_P^2}{(m_{q_j}+m_{q_i})^2},
\nonumber\\
K_b^P &=& 1 + 11\, \frac{m_P^2}{(m_{q_j}+m_{q_i})^2},
\label{VIA}
\ea
for the scalar and pseudoscalar operators,
respectively.

Although our main points do not depend on the exact values
of $K_a^P$ and $K_b^P$,
we show in Table~\ref{TAB:MesonMixing:01}
a rough estimate based on the  vacuum insertion approximation
of Eq.~\eqref{VIA} and the relevant input parameters.
%
\begin{table}[h!]
\begin{ruledtabular}
		\begin{tabular}{ccccc}
			Meson system & $K^0$--$\bar K^0$ & $B_d^0$--$\bar B_d^0$ & $B_s^0$--$\bar B_s^0$ & $D^0$--$\bar D^0$\\
\hline
			$\eta$ & $1.0$& $0.1$& $0.1$& $1.0$\\
			$\Delta m_P$ & $3.484\times 10^{-15}$& $3.337\times 10^{-13}$& $1.17\times 10^{-11}$& $6.58\times 10^{-15}$\\
			$m_P$ & $0.497$& $5.28$& $5.37$& $1.86$\\
			$f_P$ & $0.156$& $0.190$& $0.225$& $0.232$\\
			$m_{q_1},m_{q_2}$ & $m_d,m_s$& $m_d,m_b$& $m_s,m_b$& $m_u,m_c$\\
			$K_a^P$ & $25.8$& $2.58$& $2.56$& $3.12$\\
			$K_b^P$ & $273.8$& $18.3$& $18.2$& $24.3$\\
		\end{tabular}
		\caption{\label{TAB:MesonMixing:01}Meson mixing input; $\Delta m_P$,
            $M_P$, $f_P$, $m_q$ in GeV.
			Taken from Ref.~\cite{Agashe:2014kda}.}
\end{ruledtabular}
\end{table}
%
The quark masses are taken (in GeV) as
$m_u=0.0023$,
$m_d=0.0048$,
$m_s=0.095$,
$m_c=1.275$,
and $m_b=4.2$ \cite{Agashe:2014kda}.
We notice that the pseudoscalar matrix elements $K_b^p$ are
always larger than their scalar counterparts $K_a^p$.
For example, in the $B$ systems, we get
$K_b^P/K_a^P \sim 7$ in the vacuum insertion approximation.
It is interesting to compare with the values obtained from lattice.
For instances,
we find
\ba
f_{B_q}^2 K_a^{B_q}
&=&
- 5 f_{B_q}^2 B^{(2)}_{B_q} + 6 f_{B_q}^2 B^{(4)}_{B_q},
\nonumber\\
f_{B_q}^2 K_b^{B_q}
&=&
5 f_{B_q}^2 B^{(2)}_{B_q} + 6 f_{B_q}^2 B^{(4)}_{B_q},
\label{VIA_lattice}
\ea
where the right-hand side involves the quantities
introduced in Ref.~\cite{Bouchard:2011xj}\footnote{Notice that
	our conventions for matrix elements differs from those in
	Ref.~\cite{Bouchard:2011xj} by a minus sign and by
	$1/(2 m_B)$. Of course, physical results are the same and,
	when all is properly taken into account, one obtains
	 Eqs.~\eqref{VIA_lattice}.}.
In particular,
$B^{(2)}_{B_q}$ ($q=d, s$)
is the bag parameter common to the operators
${\cal O}_2 = (\bar{b} \gamma_L q)^2$
and $\tilde{{\cal O}}_2 = (\bar{b} \gamma_R q)^2$,
with $\gamma_{L,R} = (1 \mp \gamma_5)/2$,
while $B^{(4)}_{B_q}$ ($q=d, s$)
is the bag parameter of the operator
${\cal O}_4 = (\bar{b} \gamma_L q)(\bar{b} \gamma_R q)$.
The results obtained from lattice differ from those
in the vacuum insertion approximation by at most a factor of
three.
Of crucial importance is the ratio
$K_b^{B_q}/K_a^{B_q} \sim 3$ obtained from lattice.
%
%
The fact that the ratio obtained using the lattice results
of \cite{Bouchard:2011xj} is closer to unity than that obtained in the
vacuum insertion approximation will be of interest in the following.

Next we highlight one of the main points in this work.
In Eq.~\eqref{M21} the scalar and pseudoscalar components appear
independently;
there are terms in $a_{{\bar{j}}i}^2$ and $b_{{\bar{j}}i}^2$,
but no term in $a_{{\bar{j}}i} b_{{\bar{j}}i}$.
One would say that they do not interfere.
But, because $a_{{\bar{j}}i}^2$ and $b_{{\bar{j}}i}^2$ are complex,
Eq.~\eqref{M21} shows the very interesting feature that the
two terms can cancel each other.
And, as we will illustrate below,
there are generic classes of models in which they could easily
arise with the opposite sign.
This has the result that the scalar contribution of a spin zero particle
may cancel the pseudoscalar contribution of that \textit{same} spin zero particle.
There are known instances where the contribution of some scalar
cancels the contribution of some other scalar.
This occurs, for instances, in the scalar contributions to the
electric dipole moment of the electron in CP violating
two Higgs doublet models, when in the decoupling limit \cite{Jung:2013hka}.
But the feature present in Eq.~\eqref{M21} is something else.
It is the two components of a single scalar that may cancel each other.
We denote this effect by ``self-cancelation''.
As far as we know, this feature hasn't been properly appreciated before,
especially in the context of its implications for the LHC.

\subsection{\label{subsec:interfere}Self-cancelation in the
most general two Higgs doublet model}

The Yukawa interactions of quarks in the
two Higgs doublet model (2HDM)
may be written as
\be
- {\cal L}_Y = \bar{q}_L (\Gamma_1 \Phi_1 + \Gamma_2 \Phi_2) n_R
+
\bar{q}_L (\Delta_1 \tilde{\Phi}_1 + \Delta_2 \tilde{\Phi}_2) p_R + \textrm{h.c.},
\ee
where $\Phi_a$ ($a=1,2$) are the Higgs doublets,
$q_L = (p_L, n_L)^T$  is a vector in the 3-dimensional
family space of left-handed doublets,
and $n_R$ and $p_R$ are 3-dimensional vectors in the right-handed
spaces of charge $-1/3$ and $+2/3$ quarks,
respectively.
The complex $3 \times 3$ matrices
$\Gamma_1$, $\Gamma_2$, $\Delta_1$, and $\Delta_2$ contain the Yukawa couplings.

After spontaneous symmetry breaking,
the fields acquire the vacuum expectation values (vevs)
$v_1/\sqrt{2}$ and $v_2/\sqrt{2}$.
In general,
$v_1$ and $v_2$ are complex,
but one may, without loss of generality,
choose a basis where $v_1$ is real
and $v_2 = |v_2| e^{i \delta}$.
It is convenient to perform a unitary transformation into
the Higgs basis $\{H_1, H_2 \}$ through \cite{LS, BS}
\be
\left(
\begin{array}{c}
H_1\\
H_2
\end{array}
\right)
=
U^\dagger\,
\left(
\begin{array}{c}
\Phi_1\\
\Phi_2
\end{array}
\right),
\label{changeTOHiggs}
\ee
where
\be
U^\dagger
=
\frac{1}{v}\,
\textrm{diag}\left(1, e^{i \chi}\right)\,
\left(
\begin{array}{cc}
v_1^\ast & v_2^\ast \\
- v_2 & v_1
\end{array}
\right),
\label{eq:HBT}
\ee
is unitary, and $v = \sqrt{v_1^2 + v_2^2} = (\sqrt{2} G_F)^{-1/2}$.
As is clear from Eq.~\eqref{eq:HBT},
all the vev is now in $H_1$.
And, since $H_2$ has no vev,
its phase can be altered at will.
We will choose $\chi=0$,
Ref.~\cite{Ferreira:2010bm} has $\chi=-\delta+\pi$.
We may write
\ba
H_1
&=&
\left[
\begin{array}{c}
G^+\\
(v + H^\textrm{sm} + i G^0)/\sqrt{2}
\end{array}
\right],
\nonumber\\[6pt]
H_2
&=&
\left[
\begin{array}{c}
H^+\\
(R + i I)/\sqrt{2}
\end{array}
\right],
\ea
where $ G^+ $ and $ G^0 $ are the would-be Goldstone bosons
and $H^+$ is the charged scalar.
$H^\textrm{sm}$, $ R $ and $ I $ are neutral fields.
The Yukawa couplings in the Higgs basis are given by
$\Gamma_a^{H} = \Gamma_b U_{ba}$,
$\Delta_a^{H} = \Delta_b U_{ba}^\ast$.
Since only $H_1$ has a vev,
the quark masses arise solely from
$\Gamma_1^H$ and $\Delta_1^H$ as
\ba
\frac{v}{\sqrt{2}} U_{d_L}^\dagger \Gamma_1^{H} U_{d_R}
& = &
D_d = \textrm{diag} (m_d, m_s, m_b),
\nonumber\\
\frac{v}{\sqrt{2}} U_{u_L}^\dagger \Delta_1^{H} U_{u_R}
& = &
D_u = \textrm{diag} (m_u, m_c, m_t ),
\ea
where the transformations
$U_\alpha$ with $\alpha = d_L, d_R, u_L, u_R$ bring the quarks
into their mass basis.
The couplings to $H_2$ become
\ba
\frac{v}{\sqrt{2}} U_{d_L}^\dagger \Gamma_2^{H} U_{d_R}
& = &
N_d\ ,
\nonumber\\
\frac{v}{\sqrt{2}} U_{u_L}^\dagger \Delta_2^{H} U_{u_R}
& = &
N_u\ ,
\label{eq:N}
\ea
and the Yukawa lagrangian becomes
%
%
\be
- \frac{v}{\sqrt{2}} {\cal L}_Y
=
({\bar u}_L V,\ {\bar d}_L )
( D_d  H_1 + N_d H_2 )\ d_R
+\
( {\bar u}_L,\ {\bar d}_L V^\dagger )
( D_u  \tilde{H}_1 + N_u  \tilde{H}_2 )\ u_R
+\  h.c.\ ,
\label{eq:yukhiggs}
\ee
where $V = U_{u_L}^\dagger U_{d_L}$ is the CKM matrix.
In general,
$N_d$ and $N_u$ are not diagonal and,
thus,
responsible for the FCNSI involving $R$ and $I$.
Notice that $H^\textrm{sm}$ couples to the quarks proportionally
to their masses, as would the SM Higgs particle.
But, in general, $H^\textrm{sm}$ is not a mass eigenstate.

If $v_1 v_2 \neq 0$,
$H^\textrm{sm}$ and $ R $
are guaranteed not to be mass eigenstates and,
if there is CP violation in the pure scalar sector,
then $ I $ is also not a mass eigenstate.
They mix through
\be
\left(
\begin{array}{c}
H^\textrm{sm}\\
R\\
I
\end{array}
\right)
= T\,
\left(
\begin{array}{c}
S_1\\
S_2\\
S_3
\end{array}
\right),
\label{T}
\ee
into the $S_k$ ($k=1,2,3$) mass eigenstates corresponding
to the three neutral spin 0 particles.
Using Eqs.~\eqref{eq:yukhiggs} and \eqref{T},
we can finally write the Yukawa lagrangian of the neutral
scalar fields as
\be
- {\cal L}_Y
=
\sum_{k=1}^3 S_k\,
\left\{
\bar{d}\, \left[ A^{d, k} + i\, \gamma_5\,  B^{u, k} \right]\, d
+
\bar{u}\, \left[ A^{u, k} + i\,  \gamma_5\,  B^{u, k} \right]\, u
\right\},
\ee
where
\ba
v\, A^{d, k}
&=&
T_{1k}\, D_d + T_{2k}\, X_+^d + i\, T_{3k}\, X_-^d,
\nonumber\\
v\, B^{d, k}
&=&
- i\, T_{2k}\, X_-^d + T_{3k}\, X_+^d,
\nonumber\\
v\, A^{u, k}
&=&
T_{1k}\, D_u + T_{2k}\, X_+^u - i\, T_{3k}\, X_-^u,
\nonumber\\
v\, B^{u, k}
&=&
- i\, T_{2k}\, X_-^u - T_{3k}\, X_+^u,
\ea
and
\be
X_\pm^\alpha = \frac{N_\alpha \pm N_\alpha^\dagger}{2},
\ee
for $\alpha = u, d$.
The down type couplings agree with those found in
Eqs.~(22.73)-(22.74) of \cite{BLS}.
Notice that the matrices $A^{\alpha, k}$ and $B^{\alpha, k}$ are hermitian,
as needed for a hermitian lagrangian.

We may now calculate the contribution to the $P^0$--$\bar{P}^0$ mixing
matrix element $M_{21}$ in Eq.~\eqref{M21} as
\be
\frac{24}{f_P^2 m_P}\
M_{12}^k
= \frac{1}{m_{S_k}^2}
\left[
\left(
a_{{\bar j}i}^k
\right)^2
K_a^P
+
\left(
b_{{\bar j}i}^k
\right)^2
K_b^P
\right],
\ee
where
\ba
v\, a_{{\bar j}i}^k
&=&
\left(
A^{\alpha,k}
\right)_{{\bar j}i},
\nonumber\\
v\, b_{{\bar j}i}^k
&=&
\left(
B^{\alpha,k}
\right)_{{\bar j}i}
\label{M21_sc}
\ea
and $\alpha=d$ for the $K$, $B_d$, and $B_s$ systems,
while $\alpha=u$ for the $D$ system.

\subsection{\label{subsec:CPcons}Self-cancelation
in a CP conserving pure scalar sector}

Although it may seem counterintuitive,
one can have a spin 0 state which arises out of a CP conserving
Higgs potential, but which, nevertheless,
couples with quarks through both scalar and pseudoscalar components.
To illustrate this mechanism,
let us consider a model which is CP conserving in its pure scalar sector.
By this we mean that all couplings in the Higgs potential are real
and that both vevs are real.
One may write
\ba
v_1 &=& v\, c_\beta,
\nonumber\\
v_2 &=& v\, s_\beta,
\label{beta}
\ea
implying that $t_\beta = v_2/v_1$,
where,
thenceforth,
$c_\theta$, $s_\theta$, and $t_\theta$ represent the
cosine, the sine, and the tangent of some given angle
$\theta$,
respectively.
In such cases,
and continuing to consider only the scalar sector,
there is one CP odd state ($A \equiv I$),
and it is common to define the
lighter ($h$) and heavier ($H$) CP even
states by \cite{care}
\ba
\left(
\begin{array}{c}
H\\
h
\end{array}
\right)
&=&
\left(
\begin{array}{cc}
c_\alpha & s_\alpha\\
- s_\alpha & c_\alpha
\end{array}
\right)
\left(
\begin{array}{c}
\rho_1\\
\rho_2
\end{array}
\right)
\nonumber\\*[2mm]
&=&
\left(
\begin{array}{cc}
c_\alpha & s_\alpha\\
- s_\alpha & c_\alpha
\end{array}
\right)
\left(
\begin{array}{cc}
c_\beta & -s_\beta\\
- s_\beta & c_\beta
\end{array}
\right)
\left(
\begin{array}{c}
H^\textrm{sm}\\
R
\end{array}
\right)
=
\left(
\begin{array}{cc}
c_{\alpha-\beta} & s_{\alpha-\beta}\\
- s_{\alpha-\beta} & c_{\alpha-\beta}
\end{array}
\right)
\left(
\begin{array}{c}
H^\textrm{sm}\\
R
\end{array}
\right),
\ea
where $\textrm{Re}\Phi^0_a = (\rho_a + v_a)/\sqrt{2}$ ($a=1,2$),
and in going to the second line,
we have used Eqs.~\eqref{changeTOHiggs}, \eqref{eq:HBT}, and \eqref{beta}.
In this case,
Eq.~\eqref{T} becomes
\be
\left(
\begin{array}{c}
H^\textrm{sm}\\
R\\
I
\end{array}
\right)
=
\left(
\begin{array}{ccc}
c_{\alpha-\beta} & - s_{\alpha-\beta} & 0\\
s_{\alpha-\beta} & c_{\alpha-\beta} & 0\\
0 & 0 & 1
\end{array}
\right)
\,
\left(
\begin{array}{c}
H\\
h\\
A
\end{array}
\right).
\label{T_CPcons}
\ee
Thus,
the scalar and pseudoscalar couplings of each mass eigenstate
to the down type quarks become:
\ba
v\, A^{d, H}
&=&
c_{\alpha-\beta}\, D_d + s_{\alpha-\beta}\, X_+^d,
\nonumber\\
v\, B^{d, H}
&=&
- i\, s_{\alpha-\beta}\, X_-^d,
\nonumber\\
v\, A^{d, h}
&=&
- s_{\alpha-\beta}\, D_d + c_{\alpha-\beta}\, X_+^d,
\nonumber\\
v\, B^{d, h}
&=&
- i\, c_{\alpha-\beta}\, X_-^d,
\nonumber\\
v\, A^{d, A}
&=&
i\, X_-^d,
\nonumber\\
v\, B^{d, A}
&=&
X_+^d.
\ea
Similarly,
for the up type quarks we find
\ba
v\, A^{u, H}
&=&
c_{\alpha-\beta}\, D_u + s_{\alpha-\beta}\, X_+^u,
\nonumber\\
v\, B^{u, H}
&=&
- i\, s_{\alpha-\beta}\, X_-^u,
\nonumber\\
v\, A^{u, h}
&=&
- s_{\alpha-\beta}\, D_u + c_{\alpha-\beta}\, X_+^u,
\nonumber\\
v\, B^{u, h}
&=&
- i\, c_{\alpha-\beta}\, X_-^u,
\nonumber\\
v\, A^{u, A}
&=&
- i\, X_-^u,
\nonumber\\
v\, B^{u, A}
&=&
- X_+^u.
\label{AB_up}
\ea

Let us concentrate on the neutral meson systems with down-type quarks.
Since $D_d$ is diagonal,
the relevant coefficients are simplified into those
listed in Table~\ref{TAB:ab}.
%
\begin{table}[h!]
\begin{ruledtabular}
		\begin{tabular}{ccc}
			scalar & $v^2\, \left(a_{{\bar j}i}^k \right)^2$ &
			$v^2\, \left(b_{{\bar j}i}^k \right)^2$ \\
\hline
			H & $s_{\alpha-\beta}^2 \left(X_+^d \right)^2_{{\bar j}i}$ &
			$- s_{\alpha-\beta}^2 \left(X_-^d \right)^2_{{\bar j}i}$ \\
			h & $c_{\alpha-\beta}^2 \left(X_+^d \right)^2_{{\bar j}i}$ &
			$- c_{\alpha-\beta}^2 \left(X_-^d \right)^2_{{\bar j}i}$ \\
			A & $- \left(X_-^d \right)^2_{{\bar j}i}$ &
			$ \left(X_+^d \right)^2_{{\bar j}i}$ \\
		\end{tabular}
		\caption{\label{TAB:ab}Scalar and pseudoscalar couplings present in
            2HDMs with CP conservation
			in the Higgs potential and in the vevs.}
\end{ruledtabular}
\end{table}
%
Let us assume that the corresponding elements in
$X_+^d$ and $X_-^d$ have the same phase.
As we will illustrate in Section~\ref{subsec:BGL},
this is a rather common feature.
In that case,
the opposite signs appearing in the two columns of
Table~\ref{TAB:ab} imply a cancelation.
Indeed,
for each scalar particle
(for each row in Table~\ref{TAB:ab}),
the $\left(a_{{\bar j}i}^k \right)^2$
contribution has the opposite sign to the $\left(b_{{\bar j}i}^k \right)^2$
contribution.
This means that the scalar contribution of one spin 0
particle tends to cancel the pseudoscalar contribution of
the \textit{same} spin 0 particle\footnote{Notice
that this is completely unrelated to any further
cancelation which might occur between \textit{different}
spin 0 particles.
For example,
if one takes $m_H=m_h=m_A$ and
$\left(X_+^d \right)^2_{{\bar j}i} = \left(X_-^d \right)^2_{{\bar j}i}$,
then the $H$, $h$, and $A$ contributions cancel exactly.}.

In getting to Table~\ref{TAB:ab} nothing was assumed
besides CP conservation in the Higgs potential and in the vevs.
So, the self-cancelation is a generic feature of these 2HDMs.
For the self-cancelation to be complete in Eq.~\eqref{M21_sc}
one would need
\be
\left(X_+^d \right)^2_{{\bar j}i}\
K_a^P
\sim
\left(X_-^d \right)^2_{{\bar j}i}\
K_b^P
\label{cancel_Hh}
\ee
for $h$ and $H$,
while
\be
\left(X_-^d \right)^2_{{\bar j}i}\
K_a^P
\sim
\left(X_+^d \right)^2_{{\bar j}i}\
K_b^P
\label{cancel_A}
\ee
would be needed for $A$.
If all masses were of the same order,
then a cancelation in $H$ and $h$ would imply
a non-cancelation in $A$,
and vice-versa.

But there are other possibilities.
Recall that $s_{\alpha-\beta}$ controls the
coupling of $h$ to the vector bosons $ZZ$ and $W^+ W^-$.
If $h$ coincides with the scalar found a LHC,
then $s_{\alpha-\beta}$ should not differ much
from unity.
In that case,
the $c_{\alpha-\beta}^2$ factors in
Table~\ref{TAB:ab} curtail the $h$ contributions.
Then, one could have a self-cancelation in $H$ and
a small $A$ contribution due to $m_A \gg m_H$,
or vice-versa.

\subsection{\label{subsec:puzzle}A conundrum
of scalar-pseudoscalar mixing with a CP conserving Higgs potential}

In Section~\ref{subsec:CPcons} we considered models
where there is CP conservation in the Higgs potential,
which remains unbroken by the vevs.
In such cases, at tree level,
the spin 0 states are eigenstates of CP
defined in the pure scalar sector:
$H$ and $h$ are CP even,
while $A$ is CP odd.
Nevertheless,
each spin 0 particle couples to quarks as in
Eq.~\eqref{L},
meaning that it has both scalar ($a_{ji}$)
and pseudoscalar ($b_{ji}$) couplings to quarks.
Is there a contradiction?
No! There is no contradiction.

Let us start by considering the diagonal couplings.
The point is that there is no CP violation in the
pure scalar sector. This means that the parameters of the Higgs potential
are real and so are the vevs. As a result,
the tree level mass matrix for the neutral scalars is block diagonal
and there is no CP violation in the pure scalar sector.
This can be seen in a basis independent fashion through the
basis invariant measures of CPV introduced by Lavoura and Silva \cite{LS}.
They all vanish.
So, where does the CP violation in $a_{ii} b_{ii} \neq 0$
come from?
As is obvious from Eq.~\eqref{L},
it comes from the couplings with quarks;
from the $N_d$ and $N_u$ matrices in Eqs.~\eqref{eq:N},
originating in the complex Yukawa matrices and driving the
FCNSI.
Botella and Silva \cite{BS} have developed basis invariant measures of
CP violation which measure precisely this type of CP violation
arising from the beating of the scalar sector against the Yukawa sector.
And the relation between these invariants and Eq.~\eqref{L}
is discussed in sections 22.9.2-22.10 of Ref.~\cite{BLS}.
What does not seem to have been appreciate then is that such
effect can lead to self-cancelations,
thus hiding potentially interesting FCNSI phenomena.
In this respect,
we stress the results found so far.
The assumption that there are no cancelations is far from natural.
It turns out that reasonable models lead naturally to
(at least some degree of) cancelations.

Let us now look at non diagonal couplings.
Once in the scalar and quark mass basis,
the most general CP transformations can be written as \cite{BLS}
\ba
\left( {\cal CP} \right)\,
q_i\,
\left( {\cal CP} \right)^\dagger
&=&
e^{i \xi_i} \gamma^0 C {\bar{q}_i}^T,
\nonumber\\
\left( {\cal CP} \right)\,
\bar{q}_j\,
\left( {\cal CP} \right)^\dagger
&=&
- e^{- i \xi_j} q_j^T C^{-1} \gamma^0,
\nonumber\\
\left( {\cal CP} \right)\,
\Sh\,
\left( {\cal CP} \right)^\dagger
&=&
\eta_\Sh\, \Sh,
\ea
where $\xi$ are spurious phases brought about by the CP transformation \cite{BLS},
and we have considered only signs $\eta_S=\pm1$ for the scalar field.
These can be combined into
\ba
\left( {\cal CP} \right)\,
\Sh\, \bar{q}_j q_i\,
\left( {\cal CP} \right)^\dagger
&=&
\eta_\Sh
e^{i (\xi_i-\xi_j)}
\Sh\, \bar{q}_i q_j
\nonumber\\
\left( {\cal CP} \right)\,
\Sh\, \bar{q}_j \gamma_5 q_i\,
\left( {\cal CP} \right)^\dagger
&=&
- \eta_\Sh
e^{i (\xi_i-\xi_j)}
\Sh\, \bar{q}_i \gamma^5 q_j,
\ea
and Eq.~\eqref{L} is transformed into
\be
- \left( {\cal CP} \right)\, {\cal L}_Y \,
\left( {\cal CP} \right)^\dagger
=
\eta_\Sh \Sh  \left\{
e^{i (\xi_i-\xi_j)}
\bar{q}_i\, (a_{{\bar{j}}i} - i b_{{\bar{j}}i} \gamma_5)\, q_j
+
e^{i (\xi_j-\xi_i)}
\bar{q}_j\, (a_{{\bar{j}}i}^\ast - i b_{{\bar{j}}i}^\ast \gamma_5)\, q_i
\right\} + \cdots
\label{LCP}
\ee
For CP conservation to hold,
the first term of Eq.~\eqref{L} must equal the second term of
Eq.~\eqref{LCP},
leading to
\ba
a_{{\bar{j}}i}
&=&
\eta_\Sh
a_{{\bar{j}}i}^\ast
e^{i (\xi_j-\xi_i)},
\nonumber\\
b_{{\bar{j}}i}
&=&
- \eta_\Sh
b_{{\bar{j}}i}^\ast
e^{i (\xi_j-\xi_i)}.
\ea
Of course, the spurious phase $(\xi_j-\xi_i)$ can be chosen to make
either equation hold.
This is a reflection of the known fact that a term by itself cannot
lead to CP violation; one needs always the beating of two terms.
However,
these equations taken together mean that CP conservation implies
\be
a_{{\bar{j}}i} b_{{\bar{j}}i}^\ast
=
- \left(
a_{{\bar{j}}i} b_{{\bar{j}}i}^\ast
\right)^\ast,
\ee
\textit{i.e.}
$
\textrm{Re}
\left(
a_{{\bar{j}}i} b_{{\bar{j}}i}^\ast
\right)
= 0,
$
which does not depend on the spurious phases.
This is a rephasing independent sign of CP conservation.
Conversely,
\be
\textrm{Re}
\left(
a_{{\bar{j}}i} b_{{\bar{j}}i}^\ast
\right)
\neq 0
\ \
\Longrightarrow
\ \
\textrm{CP Violation}\ \ (i \neq j).
\ee
Notice the curious possibility that one could have
CP conservation with $a_{{\bar{j}}i} b_{{\bar{j}}i}^\ast \neq 0$
as long as the two couplings were relatively imaginary.
We know of no model for which this is a compulsory feature,
but the possibility should be kept in mind.
Of course,
since hermiticity of the Lagrangian requires the diagonal couplings
to be real,
\be
a_{{\bar{i}}i} b_{{\bar{i}}i}
\neq 0
\ \
\Longrightarrow
\ \
\textrm{CP Violation}\ \ (\textrm{diagonal}).
\ee
A similar analysis for the parity transformation would
lead to
\be
a_{{\bar{j}}i} b_{{\bar{j}}i}^\ast
\neq 0
\ \
\Longrightarrow
\ \
\textrm{P Violation},
\ee
regardless of $i =j$ or $i \neq j$.

\subsection{\label{subsec:BGL}Self-cancelation in BGL models}

We have mentioned that there are models where the corresponding
matrix elements of $X_+^d$ and $X_-^d$ have
a common phase.
One such example is provided by a model proposed
by Branco, Grimus, and Lavoura,
known as the BGL model \cite{BGL, Botella:2014ska}.
The model was constructed to obviate constraints on FCNSI by
relating the matrices $N_u$ or $N_d$ with off-diagonal CKM matrix elements,
which are known to be small.
As shown in Ref.~\cite{Ferreira:2010ir} under some assumptions,
BGL models provide the only possible implementation
in 2HDMs of a relation between FCNSI and the CKM matrix
which uses abelian symmetries.
There are six such models in the quark sector\footnote{These branch
into more possibilities once one takes the leptonic sector into account \cite{Botella:2011ne}.}.
Three models,
known as up models (types $u$, $c$, and $t$),
have a diagonal $N_u$ and a non diagonal $N_d$.
Three models,
known as down models (types $d$, $s$, and $b$),
have a diagonal $N_d$ and a non diagonal $N_u$.

\subsubsection{\label{subsubsec:t}Up models}

After some calculations,
we find for the type $t$ model
\begin{equation}
	N_u=
	\begin{pmatrix}
		m_{u} t_\beta & 0 & 0\\
		0& m_{c} t_\beta & 0\\
		0 & 0 & -m_{t} t_\beta^{-1}
	\end{pmatrix}
\qquad
X_+^u
=N_u,
\quad
X_-^u
=0,
\label{NU_type_t}
\end{equation}
\begin{equation}
	N_d=
	\begin{pmatrix}
		m_{d}\left[(1-|V_{td}|^2)t_\beta-|V_{td}|^2t_\beta^{-1}\right] & -m_{s}(t_\beta+t_\beta^{-1})V_{ts}V^\ast_{td} & -m_{b}(t_\beta+t_\beta^{-1})V_{tb}V^\ast_{td}\\
		-m_{d}(t_\beta+t_\beta^{-1})V_{td}V^\ast_{ts} & m_{s}\left[(1-|V_{ts}|^2)t_\beta-|V_{ts}|^2t_\beta^{-1}\right] & -m_{b}(t_\beta+t_\beta^{-1})V_{tb}V^\ast_{ts}\\
		-m_{d}(t_\beta+t_\beta^{-1})V_{td}V^\ast_{tb} & -m_{s}(t_\beta+t_\beta^{-1})V_{ts}V^\ast_{tb} & m_{b}\left[(1-|V_{tb}|^2)t_\beta-|V_{tb}|^2t_\beta^{-1}\right]
	\end{pmatrix},
\end{equation}
\begin{equation}
	X_+^d=
	\begin{pmatrix}
		m_{d}\left[(1-|V_{td}|^2)t_\beta-|V_{td}|^2t_\beta^{-1}\right] & -\frac{m_{s}+m_{d}}{2}(t_\beta+t_\beta^{-1})V_{ts}V^\ast_{td} & -\frac{m_{b}+m_{d}}{2}(t_\beta+t_\beta^{-1})V_{tb}V^\ast_{td}\\
		-\frac{m_{d}+m_{s}}{2}(t_\beta+t_\beta^{-1})V_{td}V^\ast_{ts} & m_{s}\left[(1-|V_{ts}|^2)t_\beta-|V_{ts}|^2t_\beta^{-1}\right] & -\frac{m_{b}+m_{s}}{2}(t_\beta+t_\beta^{-1})V_{tb}V^\ast_{ts}\\
		-\frac{m_{d}+m_{b}}{2}(t_\beta+t_\beta^{-1})V_{td}V^\ast_{tb} & -\frac{m_{s}+m_{b}}{2}(t_\beta+t_\beta^{-1})V_{ts}V^\ast_{tb} & m_{b}\left[(1-|V_{tb}|^2)t_\beta-|V_{tb}|^2t_\beta^{-1}\right]
	\end{pmatrix},
\end{equation}
\begin{equation}
	X_-^d=
	\begin{pmatrix}
		0 & -\frac{m_{s}-m_{d}}{2}(t_\beta+t_\beta^{-1})V_{ts}V^\ast_{td} & -\frac{m_{b}-m_{d}}{2}(t_\beta+t_\beta^{-1})V_{tb}V^\ast_{td}\\
		\frac{m_{s}-m_{d}}{2}(t_\beta+t_\beta^{-1})V_{td}V^\ast_{ts} & 0 & -\frac{m_{b}-m_{s}}{2}(t_\beta+t_\beta^{-1})V_{tb}V^\ast_{ts}\\
		\frac{m_{b}-m_{d}}{2}(t_\beta+t_\beta^{-1})V_{td}V^\ast_{tb} & \frac{m_{b}-m_{s}}{2}(t_\beta+t_\beta^{-1})V_{ts}V^\ast_{tb} & 0
	\end{pmatrix},
\end{equation}
The matrices $X_+^\alpha$ are hermitian,
while $X_-^\alpha$ are anti-hermitian.
Moreover,
as announced,
$\left(X_+^d \right)_{{\bar j}i}$
has the same phase as
$\left(X_-^d \right)_{{\bar j}i}$.
Notice that the $(t_\beta+t_\beta^{-1})$ prefactor
in the off diagonal terms can make the FCNSI large for sufficiently
large values of $\tan{\beta}$.

The matrix $N_d$ is highly hierarchical,
both due to the masses and due to the CKM matrix elements.
Concentrating on the off-diagonal matrix elements,
and taking out the $(t_\beta+t_\beta^{-1})$ prefactor,
the orders of magnitude for $X^d_+$ and $X^d_-$
in the $t$ type model are
\be
X_\pm^d \sim
\begin{pmatrix}
\textrm{x}	 & m_s\lambda^5 & m_b \lambda^3\\
 \textrm{x} & \textrm{x} & m_b \lambda^2\\
\textrm{x}	& \textrm{x}  & \textrm{x}
\end{pmatrix}
\ \ \
\textrm{($t$ type)},
\label{Xt}
\ee
where $\lambda$ is the expansion parameter in the Wolfenstein
parametrization of the CKM matrix \cite{WolfCKM}.
We show only the 12, 13, and 23 elements because
we are focusing on flavour violating transitions and
because the transposed elements are of the same order of magnitude.
Thus, in the $t$ type model,
the largest contribution would occur in the $B_s$ system.
The only difference in the $c$ type model is that the CKM combinations
$V_{\alpha  i}V^\ast_{\alpha j}$ with $\alpha=t$ get changed
into
$V_{\alpha  i}V^\ast_{\alpha j}$ with $\alpha=c$,
and similarly for the $u$ type model.
Thus,
\be
X_\pm^d \sim
\begin{pmatrix}
	\textrm{x}	 & m_s\lambda & m_b \lambda^3\\
	\textrm{x} & \textrm{x} & m_b \lambda^2\\
	\textrm{x}	& \textrm{x}  & \textrm{x}
\end{pmatrix}
\ \ \
\textrm{($c$ type)},
\label{Xc}
\ee
\be
X_\pm^d \sim
\begin{pmatrix}
	\textrm{x}	 & m_s\lambda & m_b \lambda^3\\
	\textrm{x} & \textrm{x} & m_b \lambda^4\\
	\textrm{x}	& \textrm{x}  & \textrm{x}
\end{pmatrix}
\ \ \
\textrm{($u$ type)}.
\label{Xu}
\ee
Eqs.~\eqref{Xt}-\eqref{Xu} can be used by model builders to
increase some FCNSI of interest and suppress others.

Let us focus again on the $t$ type model.
In this model
\be
\left|
\frac{X_+^d}{X_-^d}
\right|^2
=
\left(
\frac{m_{q_2} + m_{q_1}}{m_{q_2} - m_{q_1}}
\right)^2
> 1.
\ee
Since we see from Table~\ref{TAB:MesonMixing:01}
that $K_b^P > K_a^P$,
the cancelation in Eq.~\eqref{cancel_A} is not possible,
while,
due to the hierarchical mass structure of the different quark families,
the cancelation in Eq.~\eqref{cancel_Hh} is only partial,
as we see from Table~\ref{tab:ratios}.
%
\begin{table}[h!]
\begin{ruledtabular}
		\begin{tabular}{cccc}
			Meson system & $K^0$--$\bar K^0$ & $B_d^0$--$\bar B_d^0$ &
			$B_s^0$--$\bar B_s^0$\\
\hline
			$K_b^P/K_a^P$ & $10.6$& $7.1\, [2.9]$& $7.1\, [3.2]$\\
			$(m_{q_2} + m_{q_1})^2/(m_{q_2} - m_{q_1})^2$
			& $1.2$& $1.0$& $1.1$\\
		\end{tabular}
		\caption{\label{tab:ratios}Comparison of the ratios needed for
            the self-cancelation in
			Eq.~\eqref{cancel_Hh}. The numbers are obtained from
			Table~\ref{TAB:MesonMixing:01},
			except those whithin $[\ ]$, which are obtained from
			Eq.~\eqref{VIA_lattice} and the lattice
			simulation of Ref.~\cite{Bouchard:2011xj}.}
\end{ruledtabular}
\end{table}
%
Thus,
although the relative minus signs in Table~\ref{TAB:ab} indicate some
self-cancelation,
in the BGL models this cancelation is not complete
because the ratios os masses are not enough to offset the
ratios of hadronic matrix elements.
We note that the cancelation is more effective when the matrix elements are
estimated with the lattice results of Ref.~\cite{Bouchard:2011xj},
then when they are estimated in the
vacuum insertion approximation.
We conclude that this mechanism must be taken into account in the experimental search for the
features of a generic 2HDM.
Indeed, as this example shows,
an accidental cancelation is quite likely
and should not be ruled out a priori.

\subsubsection{\label{subsubsec:b}Down models}

After some calculations,
we find for the type $b$ model
\begin{equation}
	N_u=
	\begin{pmatrix}
		m_{u}\left[(1-|V_{ub}|^2)t_\beta-|V_{ub}|^2t_\beta^{-1}\right] & -m_{c}(t_\beta+t_\beta^{-1})V_{cb}V^\ast_{ub} & -m_{t}(t_\beta+t_\beta^{-1})V_{tb}V^\ast_{ub}\\
		-m_{u}(t_\beta+t_\beta^{-1})V_{ub}V^\ast_{cb} & m_{c}\left[(1-|V_{cb}|^2)t_\beta-|V_{cb}|^2t_\beta^{-1}\right] & -m_{t}(t_\beta+t_\beta^{-1})V_{tb}V^\ast_{cb}\\
		-m_{u}(t_\beta+t_\beta^{-1})V_{ub}V^\ast_{tb} & -m_{c}(t_\beta+t_\beta^{-1})V_{cb}V^\ast_{tb} & m_{t}\left[(1-|V_{tb}|^2)t_\beta-|V_{tb}|^2t_\beta^{-1}\right]
	\end{pmatrix}
\end{equation}
\begin{equation}
	X_+^u =
	\begin{pmatrix}
		m_{u}\left[(1-|V_{ub}|^2)t_\beta-|V_{ub}|^2t_\beta^{-1}\right] & -\frac{m_{u}+m_{c}}{2}(t_\beta+t_\beta^{-1})V_{cb}V^\ast_{ub} & -\frac{m_{u}+m_{t}}{2}(t_\beta+t_\beta^{-1})V_{tb}V^\ast_{ub}\\
		-\frac{m_{u}+m_{c}}{2}(t_\beta+t_\beta^{-1})V_{ub}V^\ast_{cb} & m_{c}\left[(1-|V_{cb}|^2)t_\beta-|V_{cb}|^2t_\beta^{-1}\right] & -\frac{m_{c}+m_{t}}{2}(t_\beta+t_\beta^{-1})V_{tb}V^\ast_{cb}\\
		-\frac{m_{u}+m_{t}}{2}(t_\beta+t_\beta^{-1})V_{ub}V^\ast_{tb} & -\frac{m_{c}+m_{t}}{2}(t_\beta+t_\beta^{-1})V_{cb}V^\ast_{tb} & m_{t}\left[(1-|V_{tb}|^2)t_\beta-|V_{tb}|^2t_\beta^{-1}\right]
	\end{pmatrix}
\end{equation}
\begin{equation}
	X_-^u=
	\begin{pmatrix}
		0 & -\frac{m_{c}-m_{u}}{2}(t_\beta+t_\beta^{-1})V_{cb}V^\ast_{ub} & -\frac{m_{t}-m_{u}}{2}(t_\beta+t_\beta^{-1})V_{tb}V^\ast_{ub}\\
		\frac{m_{c}-m_{u}}{2}(t_\beta+t_\beta^{-1})V_{ub}V^\ast_{cb} & 0 & -\frac{m_{t}-m_{c}}{2}(t_\beta+t_\beta^{-1})V_{tb}V^\ast_{cb}\\
		\frac{m_{t}-m_{u}}{2}(t_\beta+t_\beta^{-1})V_{ub}V^\ast_{tb} & \frac{m_{t}-m_{c}}{2}(t_\beta+t_\beta^{-1})V_{cb}V^\ast_{tb} & 0
	\end{pmatrix}
\end{equation}
\begin{equation}
	N_d=
	\begin{pmatrix}
		m_dt_\beta & 0 & 0\\
		0& m_st_\beta & 0\\
		0 & 0 & -m_bt_\beta^{-1}
	\end{pmatrix},
\qquad
X_+^d=
N_d,
\quad
X_-^d
=0.
\end{equation}

In terms of orders of magnitude,
we have
\be
X_\pm^u \sim
\begin{pmatrix}
	\textrm{x}	 & m_c \lambda^5 & m_t \lambda^3\\
	\textrm{x} & \textrm{x} & m_t \lambda^2\\
	\textrm{x}	& \textrm{x}  & \textrm{x}
\end{pmatrix}
\ \ \
\textrm{($b$ type)},
\label{Xb}
\ee
\be
X_\pm^u \sim
\begin{pmatrix}
	\textrm{x}	 & m_c \lambda & m_t \lambda^3\\
	\textrm{x} & \textrm{x} & m_t \lambda^2\\
	\textrm{x}	& \textrm{x}  & \textrm{x}
\end{pmatrix}
\ \ \
\textrm{($s$ type)},
\label{Xs}
\ee
\be
X_\pm^u \sim
\begin{pmatrix}
	\textrm{x}	 & m_c \lambda & m_t \lambda^3\\
	\textrm{x} & \textrm{x} & m_t \lambda^4\\
	\textrm{x}	& \textrm{x}  & \textrm{x}
\end{pmatrix}
\ \ \
\textrm{($d$ type)}.
\label{Xd}
\ee
Notice that the orders of magnitude of the CKM coefficients in
Eqs.~\eqref{Xb}-\eqref{Xd} reproduce those in
Eqs.~\eqref{Xt}-\eqref{Xu},
respectively;
the masses are, obviously, different.
Again,
this is not enough to make the self-cancelation
fully effective for BGL models in the $D$ system.

\section{\label{sec:prod}FCNSI-induced Higgs production at the LHC}

The Lagrangian in Eq.~\eqref{L} induces $pp \rightarrow \Sh $
direct production with $\bar{q}_j$ and $q_i$
as partons (and also, with partons $\bar{q}_i$ and $q_j$).
At leading order in the narrow width approximation,
we find
\be
\sigma\left[p p (\bar{q}_j q_i) \rightarrow \Sh ) \right]
=
2\,
\frac{\pi}{8 N_c E^2}
\left(
|a_{{\bar{j}}i}|^2 + |b_{{\bar{j}}i}|^2
\right)
I_{\bar{q}_j q_i}\, ,
\label{sigma}
\ee
where
\be
I_{\bar{q}_j q_i}
=
\int_{x_0}^1 dx f_{\bar{q}_j}^p(x,Q^2)  f_{q_i}^p(x_0/x,Q^2) \frac{1}{x}\, .
\ee
A factor of two was explicitly included in Eq.~\eqref{sigma}
(due to the fact that a given parton can come with equal probability from either
proton -- things would be different in a $p \bar{p}$ collision),
$N_c$ is the number of colours,
$E$ is the energy of the colliding proton,
$f_X^p$ are the relevant parton distribution functions (PDFs),
and
\be
x_0 = \frac{m_\Sh ^2}{4 E^2}.
\ee
Notice that
\be
\frac{\sigma\left[p p (\bar{q}_i q_j) \rightarrow \Sh )
	\right]}{\sigma\left[p p (\bar{q}_j q_i) \rightarrow \Sh ) \right]}
=
\frac{I_{\bar{q}_i q_j}}{I_{\bar{q}_j q_i}}
\ee
is not unity, since, in general
$f_{\bar{q}_i}^p \neq f_{\bar{q}_j}^p$
and
$ f_{q_j}^p \neq  f_{q_i}^p$.
For instances,
although Eq.~\eqref{hermiticity} implies that
$|a_{\bar{c} u}| =|a_{\bar{u} c}|$
and
$|b_{\bar{c} u}| =|b_{\bar{u} c}|$,
$\sigma\left[p p (\bar{c} u) \rightarrow \Sh) \right]$
does not equal
$\sigma\left[p p (\bar{u} c) \rightarrow \Sh) \right]$.

\begin{figure}[h!]
	\centering
	\includegraphics[width=0.4\linewidth]{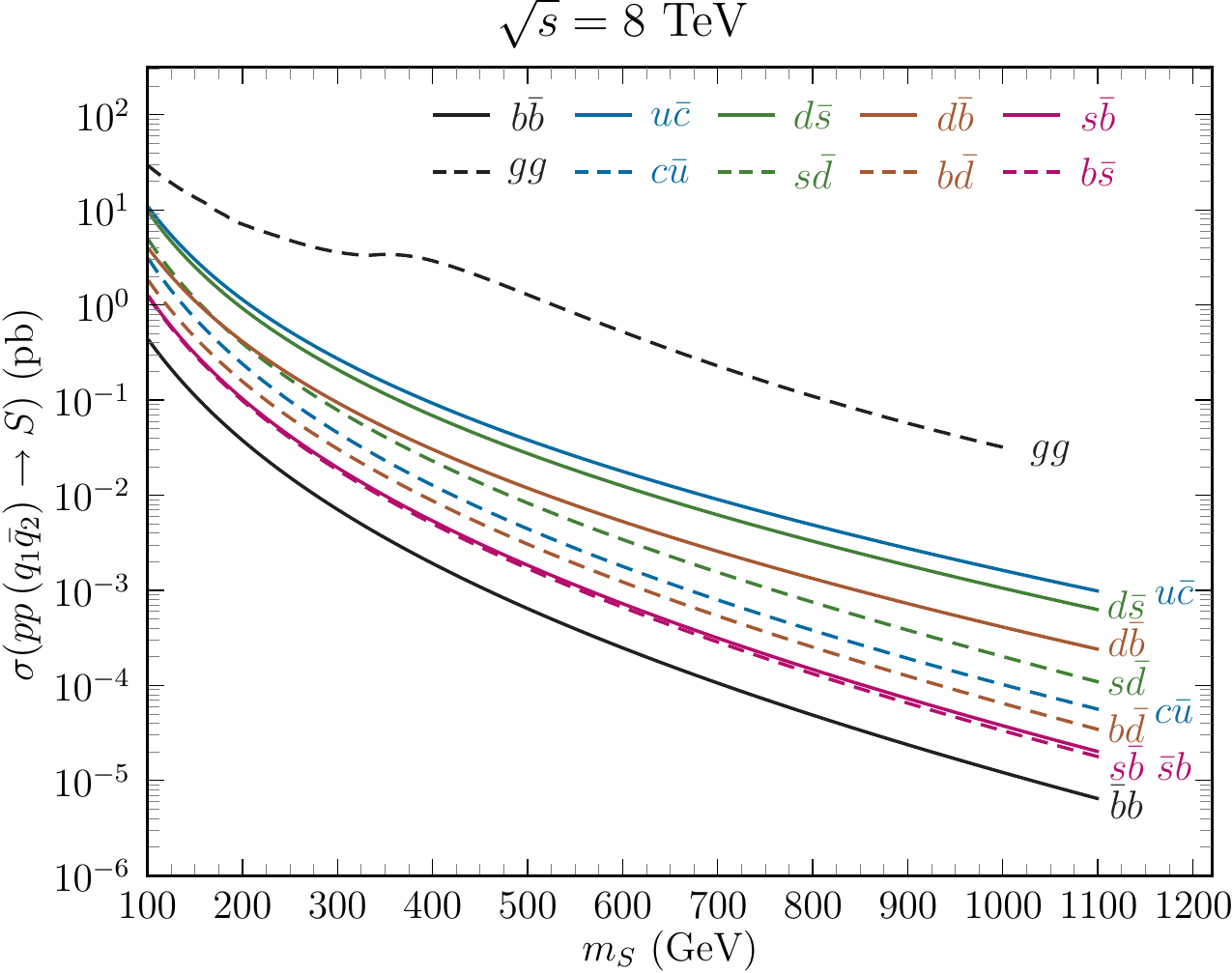}
	\hspace{0.02\linewidth}
	\includegraphics[width=0.4\linewidth]{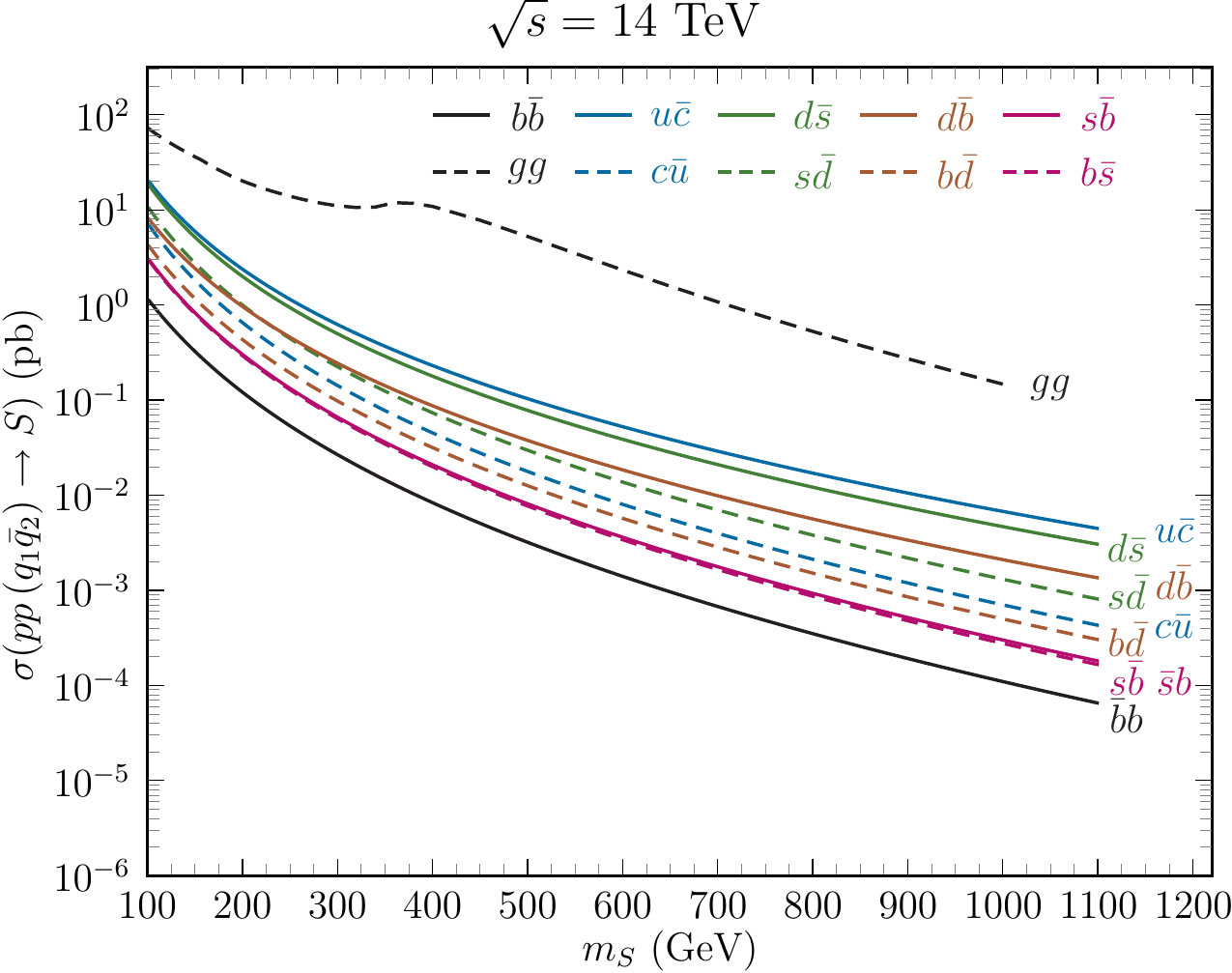}
	\caption{Cross section for $\Sh $ production in $pp$ collisions
		at 8TeV-Left panel (14TeV-Right panel) through the
		$q_i q_j$ partons indicated, assuming that they couple as the
		SM $b\bar{b}$. For reference, we show also the cross-section
		for glue-glue production.}
	\label{fig:1}
\end{figure}

In order to estimate the change obtained in going from LO to NNLO,
we use SusHi (version 1.5.0) \cite{Harlander:2012pb}
with the factorization and running scales
$\mu_F = \mu_R = m_S/4$ \cite{Dicus:1988cx, Balazs:1998sb,
Maltoni:2003pn, Harlander:2003ai, Harlander:2011aa,
Dittmaier:2011ti, Dittmaier:2012vm, website} in order
to find
\be
R(m_S) = \frac{\sigma\left[p p (\bar{b} b) \rightarrow \Sh)
	\right]_\textrm{NNLO}^\textrm{SusHi}}{
\sigma\left[p p (\bar{b} b) \rightarrow \Sh) \right]_\textrm{LO}}.
\label{R_mS}
\ee
For each scalar mass $m_S$,
we use $R(m_S)$ as a universal rescaling factor for all our
production cross sections,
thus taking into account the major factors appearing in going from
LO to NNLO.
%
%
To see the effect of the different PDFs,
we show in Fig.~\ref{fig:1} the cross sections calculated
as a function of $m_\Sh $, assuming that all
couplings coincide with the
SM $b\bar{b}$ couplings $a_{{\bar{b}}b} = m_b/v$ and
$b_{{\bar{b}}b} = 0$.
%
%
Also shown is the SM gluon-gluon fusion production cross section,
typically two to three orders of magnitude larger than the SM $b\bar{b}$.
Of course, as seen in Fig.~\ref{fig:1},
if the FCNSI couplings were all equal to $m_b/v$,
then the fact that the $u$ and $d$ PDFs in the proton are larger
than all others,
would mean that the productions through $u \bar{c}$, $d \bar{s}$, and
$d \bar{b}$ would be the largest.
We see also that there is a very small difference between
$s \bar{b}$ and $b \bar{s}$.

Since the bounds from leptonic decays depend also on the
details of the leptonic sector,
we will concentrate on the bounds from
$P^0 - \overline{P^0}$ mixing.
Let us decide that the scalar contribution to $|M_{21}|$ in
Eq.~\eqref{M21} is a fraction $\eta$ of the total contribution,
including also the SM box diagram:
\be
\left| M_{12}^\Sh  \right| = \eta \left| M_{12} \right|.
\label{M12_eta}
\ee
In the notation of the CKMfitter group \cite{ckmfitter,Lenz:2012az},
\be
M_{12}^\textrm{sm} + M_{12}^\Sh  = M_{12} = M_{12}^\textrm{sm}\, \Delta.
\ee
Thus
\be
\eta = \left| \frac{\Delta - 1}{\Delta} \right|.
\ee
In the notation of the UTfit collaboration,
$\Delta = C e^{2 i \phi}$ \cite{Bona:2007vi}.
For the $B$ systems,
a very conservative guess would be $10\%$.
In the $K$ and $D$ systems,
the long distance contributions have a large uncertainty which
could easily hide a scalar contributions amounting
to $200\%$ of the SM contributions, but with the opposite sign.

\subsection{\label{subsec:pure}Pure scalar or pseudoscalar couplings}

We start by assuming that the scalar $\Sh$ is either a pure scalar ($b=0$)
or a pure pseudoscalar ($a=0$).
From Eqs.~\eqref{M21}, \eqref{DeltaM}, and \eqref{M12_eta},
we find the upper limit
\be
\left| c^\textrm{max} \right|=
\sqrt{\frac{12}{K_c^P} \frac{\eta\, \Delta m_P}{m_P}}\ \frac{m_\Sh }{f_P},
\label{cmax}
\ee
where $c=a, b$,
and which depends linearly on $m_\Sh$,
as announced in Eq.~\eqref{lin}.

We have mentioned in connection with Table~\ref{TAB:MesonMixing:01}
and Eq.~\eqref{VIA} that
the pseudoscalar matrix elements $K_b^p$ are
always larger than their scalar counterparts $K_a^p$
(at least in the vacuum insertion approximation and the lattice
estimate of Ref.~\cite{Bouchard:2011xj} used here).
This means that the maximum allowed values for the pseudoscalar
couplings ($b$) will always be smaller than the
corresponding scalar couplings ($a$) by a factor
of roughly $3$ ($\sqrt{3}$ using Ref.~\cite{Bouchard:2011xj}).
Of course,
since we are using estimates of these matrix elements,
all results must be taken as indicative rather than
tight constraints.
For $m_h=125\textrm{GeV}$ ($m_h=1\textrm{TeV}$),
we find the maximum values
shown in Table~\ref{TAB:MesonMixing:03}.
%
\begin{table}[h!]
\begin{ruledtabular}
		\begin{tabular}{ccccc}
			Meson system & $K^0$--$\bar K^0$ & $B_d^0$--$\bar B_d^0$ &
			$B_s^0$--$\bar B_s^0$ & $D^0$--$\bar D^0$\\
\hline
			$\left| a^\textrm{max} \right|\ (125\textrm{GeV})$ &
			$4.58\times 10^{-5}$& $1.13\times 10^{-4}$&
			$5.61\times 10^{-4}$& $6.28\times 10^{-5}$\\
			$\left| b^\textrm{max} \right|\  (125\textrm{GeV})$ &
			$1.41\times 10^{-5}$& $4.24\times 10^{-5}$&
			$2.11\times 10^{-4}$& $2.25\times 10^{-5}$\\
			$\left| a^\textrm{max} \right|\  (1\textrm{TeV})$ &
			$3.66\times 10^{-4}$& $9.02\times 10^{-4}$&
			$4.49\times 10^{-3}$& $5.03\times 10^{-4}$\\
			$\left| b^\textrm{max} \right|\  (1\textrm{TeV})$ &
			$1.12\times 10^{-4}$& $3.39\times 10^{-4}$&
			$1.68\times 10^{-3}$& $1.80\times 10^{-4}$\\
		\end{tabular}
		\caption{\label{TAB:MesonMixing:03}Meson mixing constraints on
            couplings for $m_\Sh=125$ GeV and for $m_\Sh=1$ TeV.
			For comparison, the $b\bar b \Sh$ coupling is
			$\frac{m_b}{v}=1.138\times 10^{-2}$,
			for a running mass of $m_b=2.8 $ GeV.}		
\end{ruledtabular}
\end{table}
%
For comparison, the $b\bar bh$ coupling is
$c_{\Sh bb}^\textrm{sm}=\frac{m_b}{v}=1.138\times 10^{-2}$,
for a running mass of $m_b=2.8 $ GeV.
The best case occurs in the $B_s^0$--$\bar B_s^0$ system with scalar coupling,
but still
\be
\frac{ \left| a^\textrm{max}_{bs} \right|}{c_{\Sh bb}^\textrm{sm}}
= 0.05\ \ (125\textrm{GeV}),
\ \ \ \ \
\frac{ \left| a^\textrm{max}_{bs} \right|}{c_{\Sh bb}^\textrm{sm}}
= 0.4\ \ (1\textrm{TeV}).
\ee

\begin{figure}[h!]
	\centering
	\includegraphics[width=0.4\linewidth]{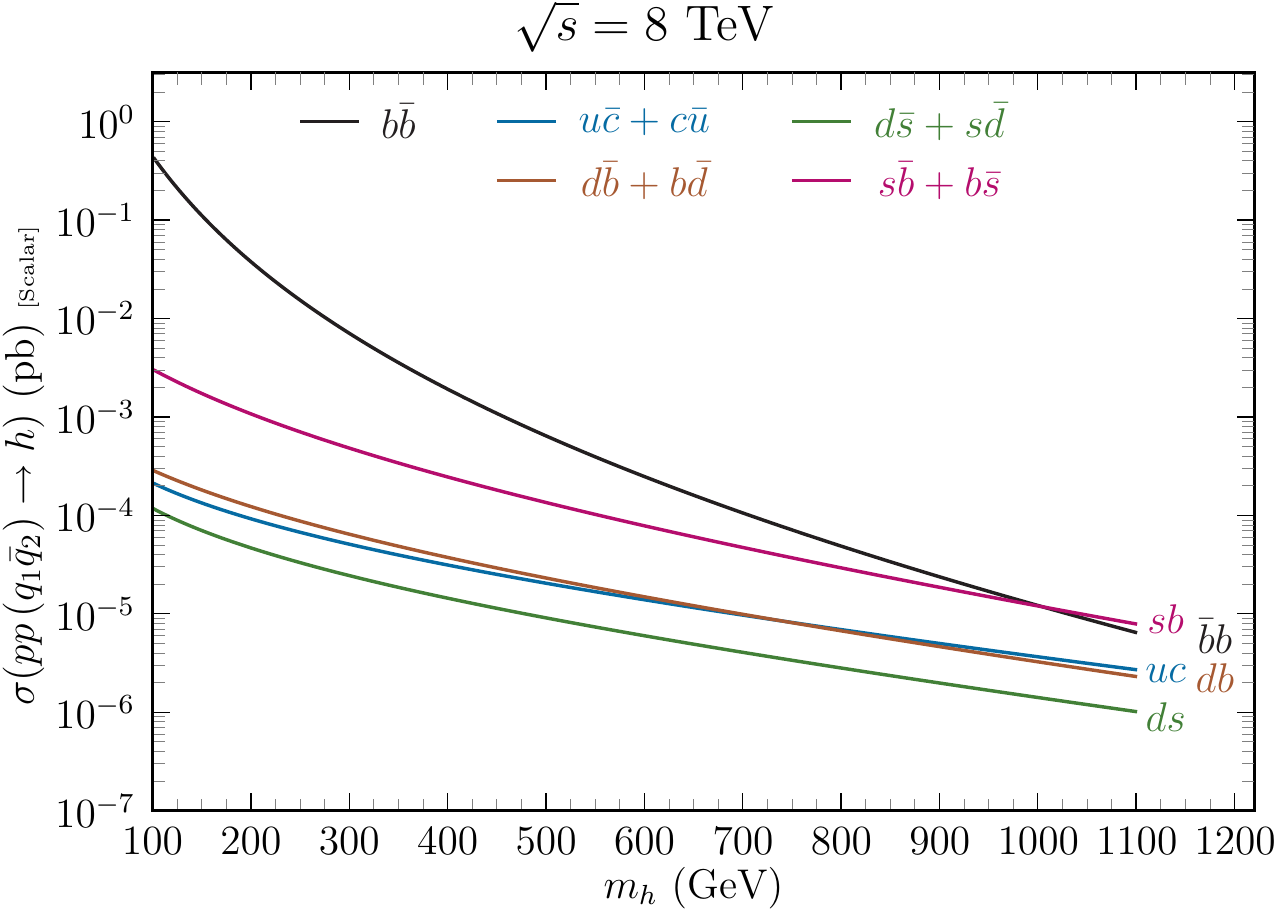}
	\hspace{0.02\linewidth}
	\includegraphics[width=0.4\linewidth]{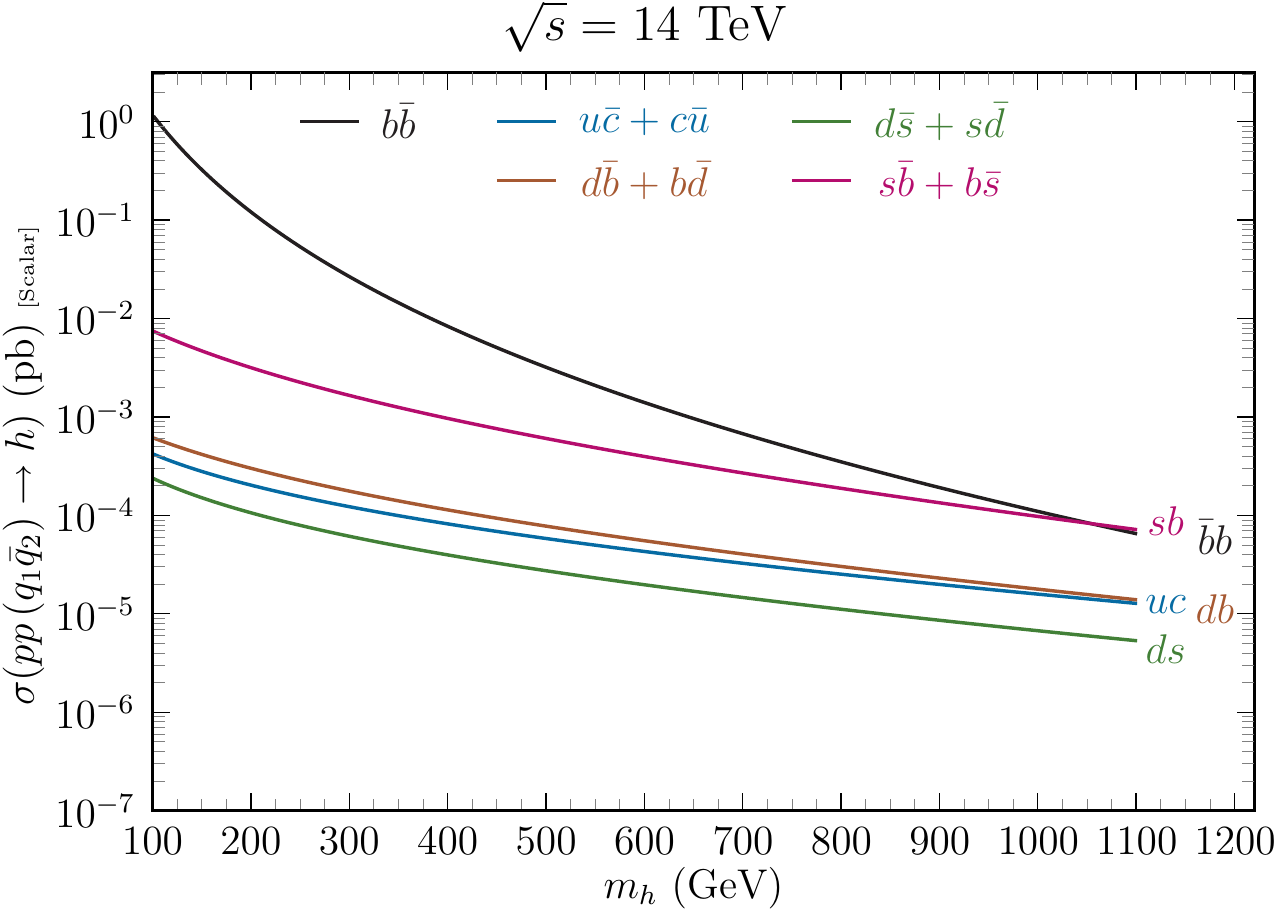}
	\caption{Sum of the cross sections
		$\sigma(pp\, (q_1\bar q_2)\to \Sh )$ + $\sigma(pp\, (q_2\bar q_1)\to \Sh )$
		for the maximal values of the pure scalar couplings consistent
		with meson mixings, for 8TeV-Left panel (14TeV-Right panel).}
	\label{fig:2}
\end{figure}

\begin{figure}[h!]
	\centering
	\includegraphics[width=0.4\linewidth]{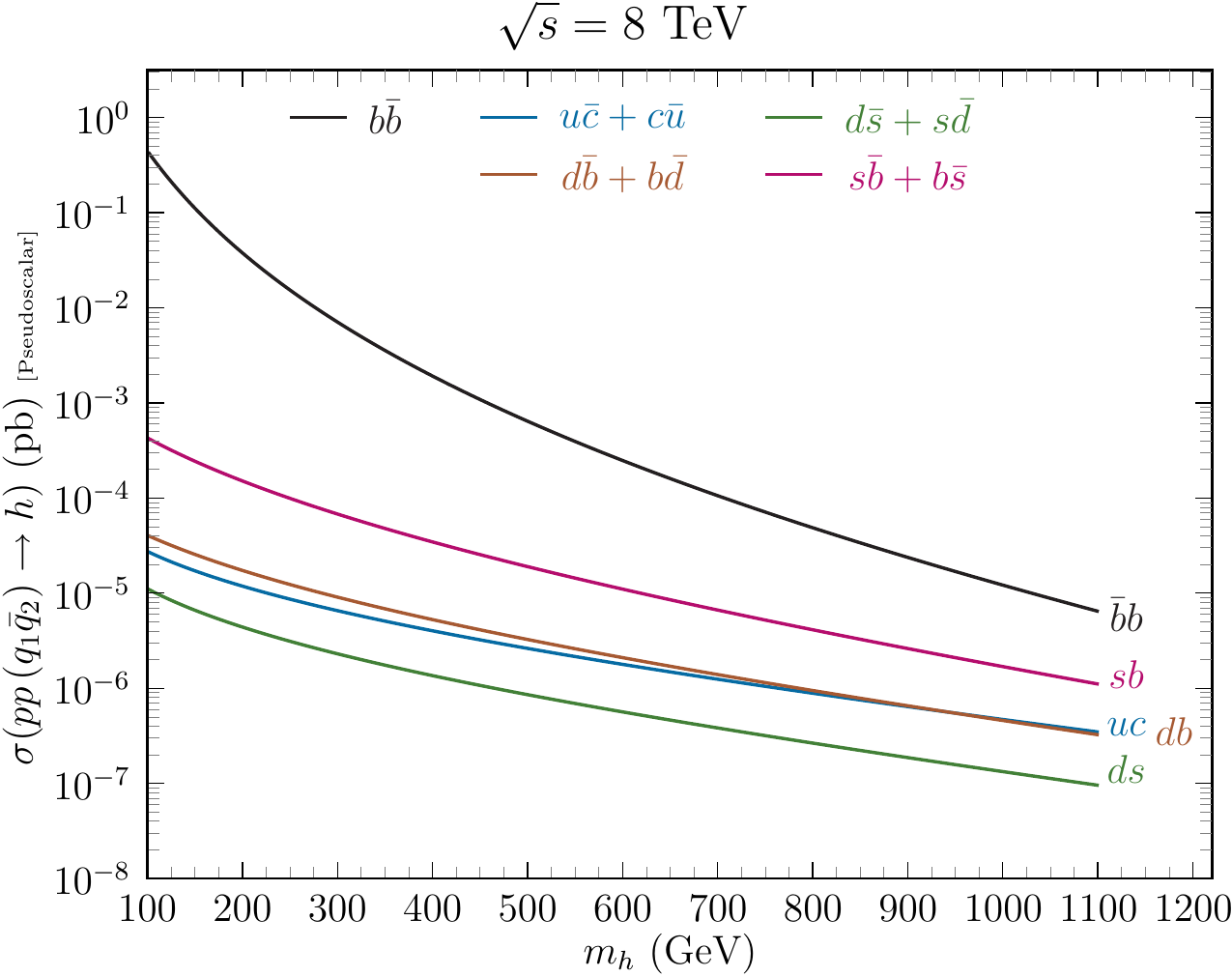}
	\hspace{0.02\linewidth}
	\includegraphics[width=0.4\linewidth]{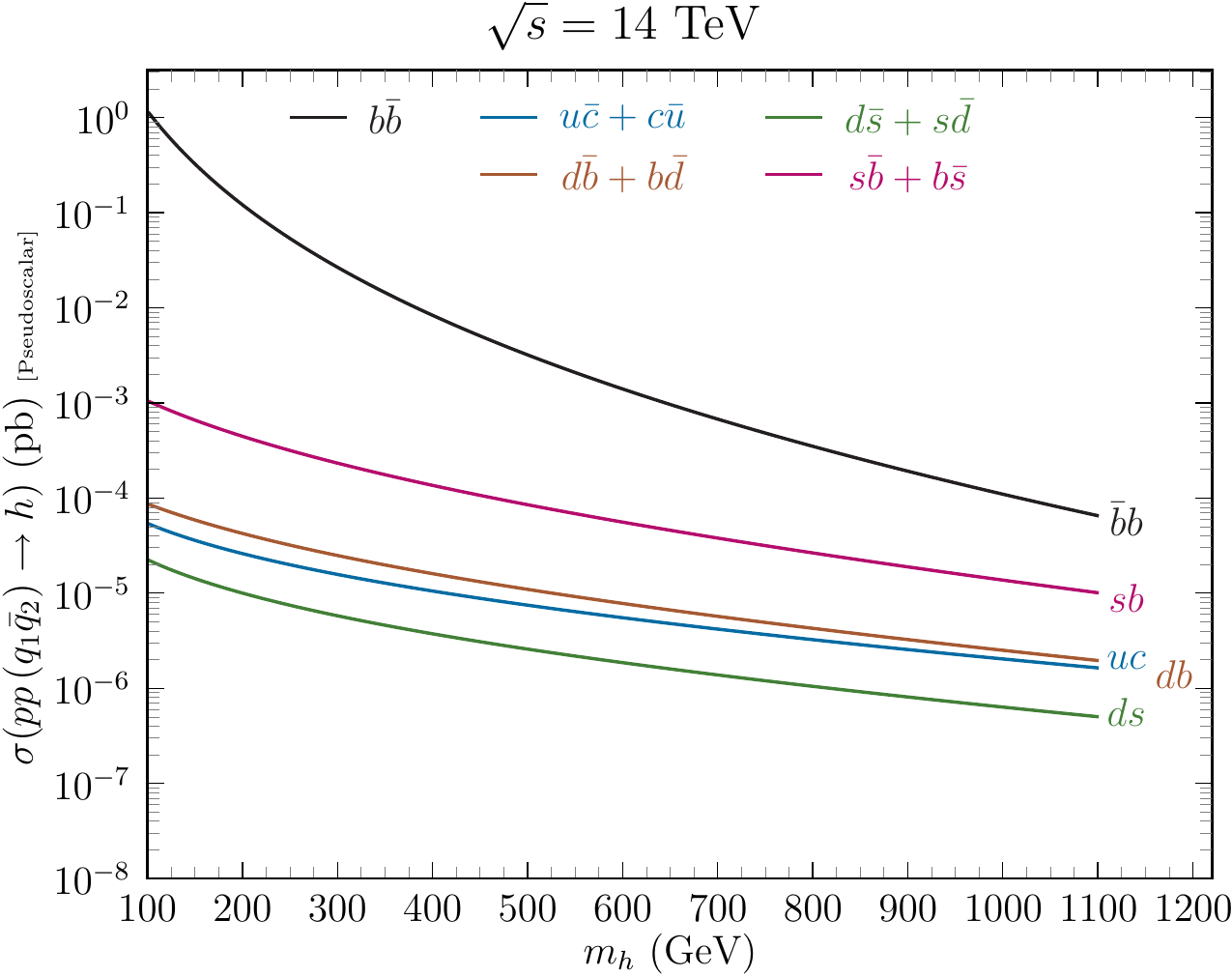}
	\caption{Sum of the cross sections
		$\sigma(pp\, (q_1\bar q_2)\to \Sh )$ + $\sigma(pp\, (q_2\bar q_1)\to \Sh )$
		for the maximal values of the pure pseudoscalar couplings consistent
		with meson mixings, for 8TeV-Left panel (14TeV-Right panel).}
	\label{fig:3}
\end{figure}

Fig.~\ref{fig:2} shows the cross sections
for production through $q_1 \bar{q}_2$  and $q_2 \bar{q}_1$
obtained when the couplings are pure scalar ($b=0$),
having the largest possible magnitude consistent with
meson mixings.
Recall that the value for $\left| a^\textrm{max} \right|$
increases with $m_S$.
%
%
Similarly,
Fig.~\ref{fig:3} shows the cross sections
for production through $q_1 \bar{q}_2$  and $q_2 \bar{q}_1$
obtained when the couplings are pure pseudoscalar ($a=0$),
having the largest possible magnitude consistent with
meson mixings.
Now one needs the value for $\left| b^\textrm{max} \right|$,
which also increases with $m_S$.
%

We stress that our concern here is not on exact values.
Our values for $\eta$ and the hadronic matrix elements are taken just
to illustrate the order of the effects.
As we will see in the nest section,
the mechanism of self-cancelation can enhance the allowed
effects and, thus,
it should be taken as a possibility in phenomenological
searches.

\subsection{\label{subsec:self}Self-cancelations and fine tuning}

We now inquire how much fine tuning might one need in
the self-cancelation,
in order that $\sigma\left[p p (\bar{q}_j q_i) \rightarrow \Sh ) \right]$
might be a relevant portion of
the production mechanism.
For definiteness,
we concentrate on the $B_s$ system.
For each value of $m_S$,
we define
\ba
r
&=&
\frac{\sigma\left[p p (\bar{b} s + \bar{s} b) \rightarrow \Sh )
	\right]_\textrm{NNLO}}{
	\sigma\left[p p (gg) \rightarrow \Sh ) \right]_\textrm{SM}}
= \frac{R\, \sigma\left[p p (\bar{b} s + \bar{s} b) \rightarrow \Sh )
	\right]_\textrm{LO}}{
	\sigma\left[p p (gg) \rightarrow \Sh ) \right]_\textrm{SM}}
\nonumber\\
&=&
\frac{\pi R}{4 N_c E^2}
\left(
|a_{{\bar{b}}s}|^2 + |b_{{\bar{b}}s}|^2
\right)
\frac{I_{\bar{b} s} + I_{\bar{s} b} }{
	\sigma\left[p p (gg) \rightarrow \Sh ) \right]_\textrm{SM}}.
\label{r}
\ea
In many circumstances
$\sigma\left[p p (gg) \rightarrow \Sh ) \right]_\textrm{SM}$
is not an appropriate reference value,
because the coupling to the top quark
(needed for the top quark triangle diagram driving the
production through gluon-gluon fusion in the SM)
is suppressed for extra scalars.
As an example,
let us consider the type $t$ model of Eq.~\eqref{NU_type_t}.
We have $X_- = 0$ and the $(X_+)_{tt}$ coupling is suppressed by
$t_\beta^{-1}$ for large $t_\beta$.
From Eqs.~\eqref{AB_up} we see that the first piece in the $A^{u, H}_{tt}$ coupling
is proportional to $m_t c_{\alpha-\beta}$.
The current LHC bounds on decays into vector bosons
constrain $s_{\alpha-\beta}^2$ to lie within $20\%$
of the SM value $1$. Assuming the central value remains unity,
Run 2 at the LHC should improve this bound to within $5\%$.
As a result,
the gluon-gluon fusion production of $H$ would be down
with respect to the SM by roughly $0.05$.
Under these circumstances,
the natural reference value for the gluon-gluon fusion production
value would be $5\%$ of the SM value.
This can be accounted for by taking as the natural value $r=0.05$.

\begin{figure}[h!]
	\centering
	\includegraphics[width=0.4\linewidth]{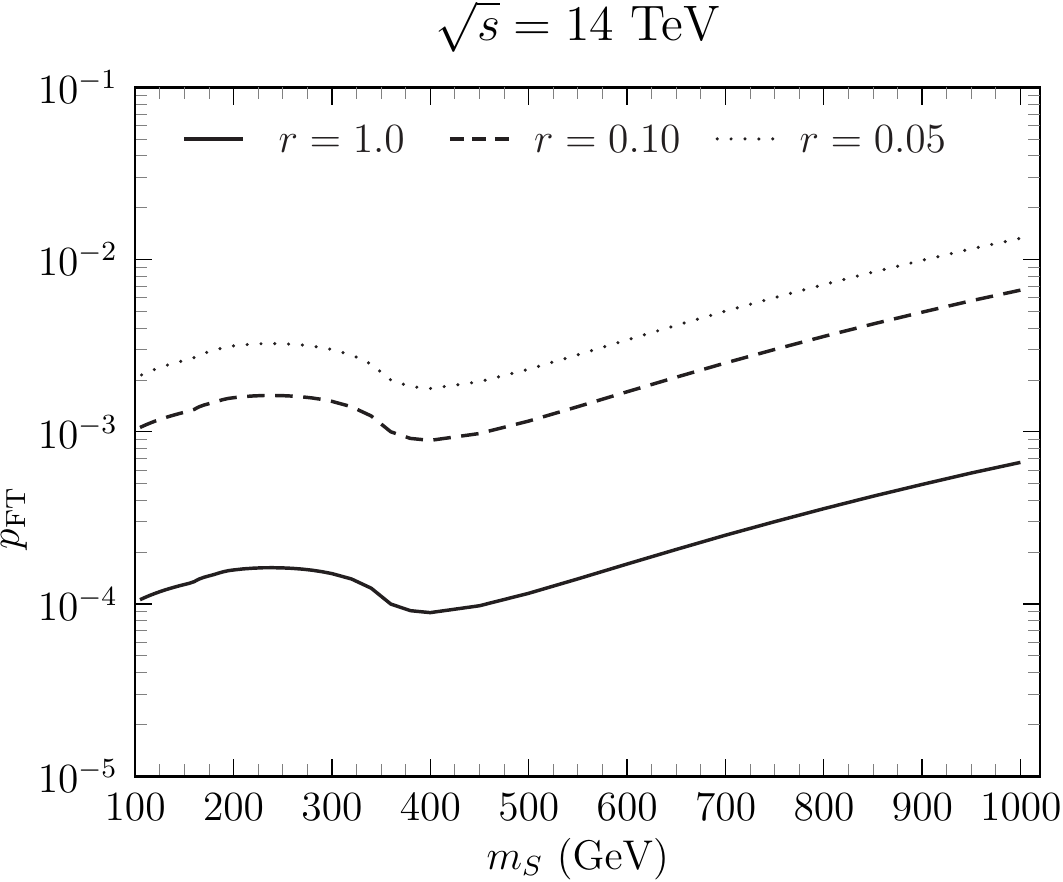}
	\caption{$p_\textrm{FT}$ as a function of
		$m_S$ for $r=1$ (solid line),
		$r=0.1$ (dashed), and $r=0.05$ (dotted).}
	\label{fig:4}
\end{figure}

For a given value of $r$ (say, 10\%),
Eq.~\eqref{r} gives a limit on
%
$|a^\textrm{max}_{{\bar{b}}s}|^2 + |b^\textrm{max}_{{\bar{b}}s}|^2$.
Of course,
these values will be much larger than those obtained from
Eq.~\eqref{cmax}.
We would now like these values of $|a^\textrm{max}_{{\bar{b}}s}|$
and $ |b^\textrm{max}_{{\bar{b}}s}|$ to survive the
$\Delta m_{B_s}$ bound through the self-cancelation mechanism.
We define the proportion of fine tuning required by
\ba
p_\textrm{FT}
&=&
\frac{|a_{{\bar{b}}s}^2 + b_{{\bar{b}}s}^2 \, K_b^{B_s}/ K_a^{B_s}|
	}{|a_{{\bar{b}}s}|^2 + |b_{{\bar{b}}s}|^2}
\nonumber\\
&=&
\frac{\eta\, \Delta m_{B_s}}{K_a^{B_s}\, m_{B_s}}
\left( \frac{m_S}{f_{B_s}}\right)^2
\frac{\pi R \left(I_{\bar{b} s} + I_{\bar{s} b}\right)}{
	E^2\, r \sigma\left[p p (gg) \rightarrow \Sh ) \right]_\textrm{SM} },
\ea
where we have used Eqs.~\eqref{M21}, \eqref{DeltaM}, \eqref{M12_eta},
and \eqref{r}.
For example,
$p_\textrm{FT}=0.01$ implies a fine tuning within $1\%$.
A larger value of $p_\textrm{FT}$ implies a lower fine tuning.
Of course,
the smaller the value chosen for $r$,
the smaller the fine tuning (larger $p_\textrm{FT}$).
This is clearly seen in Fig.~\ref{fig:4},
showing $p_\textrm{FT}$ as a function of
$m_S$ for $r=1$,
$r=0.1$, and $r=0.05$.
%
%
Consider, for example,
a situation similar to that in the type $t$ model,
where the natural gluon-gluon production corresponds to $r=0.05$,
and take $m_S=800 \textrm{GeV}$.
We learn from Fig.~\ref{fig:4} that for $bs$ production to
be equal to the gluon-gluon production requires a fine tuning
in the self-cancelation within $1\%$.
But this means that a reasonable self-cancelation within $10\%$,
would imply a $bs$ production to contribute around $10\%$ of the
total production.

Notice that, in a generic model, the couplings of the spin 0 particle to $bs$,
$bd$, and $ds$ are independent.
Thus, in a model independent search,
each neutral meson system should be considered independently.
However,
in many models the couplings in each sector could have specific relations,
as illustrated above in the BGL system.
For definiteness,
let us assume that the scalar $S$ couples like one of the scalars in the type $t$ model.
According to Table~\ref{TAB:ab}, the couplings go like
$\left(X_\pm^d \right)^2_{{\bar j}i}$, which scale like the square of the
entries of the matrix in Eq.~\eqref{Xt}.
Namely,
for the type $t$ model\footnote{Notice that, in the type $t$ mode,
the CKM coefficients in Eqs.~\eqref{scale} coincide with the
coefficients that arise from the respective SM box diagrams.
This means that the CKM multiplicative factor in the new contributions scale from
system to system in accordance with experiment.}:
\ba
K &:&
m_s^2\, V_{ts}^2 V_{td}^{2\, \ast},
\nonumber\\
B_d &:&
m_b^2\, V_{tb}^2 V_{td}^{2\, \ast},
\nonumber\\
B_s &:&
m_b^2\, V_{tb}^2 V_{ts}^{2\, \ast}.
\label{scale}
\ea
The key point is that both the scalar ($a$) and
pseudoscalar ($b$) coefficients scale in the same way.
From Eqs.~\eqref{M21}-\eqref{DeltaM},
we get
\be
\frac{\Delta m_{B_s}}{\Delta m_{B_d}}
=
\frac{f_{B_s}^2\, m_{B_s}}{f_{B_d}^2\, m_{B_d}}
\frac{\left[
a_{{\bar{b}}s}^2 K_a^{B_s} + b_{{\bar{b}}s}^2 K_b^{B_s}
\right]}{\left[
a_{{\bar{b}}d}^2 K_a^{B_d} + b_{{\bar{b}}d}^2 K_b^{B_d}
\right]
}.
\ee
Since $K_b/K_a$ is almost the same in the $B_d$ and $B_s$ systems
and since, for each system, $a$ and $b$ scale in the same way,
we expect that a 10\% self-cancelation in the $B_s$ system
will also imply a 10\% self-cancelation in the $B_d$ system.
For the kaon system,
besides the SM-like CKM rescaling,
there is an $m_s^2/m_b^2$ suppression.
Indeed,
taking
\be
c_{{\bar{s}}d} \sim \frac{m_s^2\, \lambda^{10}}{m_b^2\, \lambda^4} c_{{\bar{b}}s}
\ee
for $c=a,b$, we find,
\be
\Delta m_K =
\left(\frac{m_s}{m_b} \lambda^3 \right)^2
\frac{\left[
a_{{\bar{b}}s}^2 K_a^{K} + b_{{\bar{b}}s}^2 K_b^{K}
\right]}{\left[
a_{{\bar{b}}s}^2 K_a^{B_s} + b_{{\bar{b}}s}^2 K_b^{B_s}
\right]}\,
\frac{f_K^2\, m_K}{f_{B_s}^2\, m_{B_s}}\,
\Delta m_{B_s}.
\ee
This means that a 10\% self-cancelation in $B_s$ implies
very small contributions to the mixing in the kaon system.

Should it turn out that the self-cancelation mechanism is particularly effective
(such that the FCNSI induced scalar production is a relevant percentage of the
with gluon-gluon production),
then one must look to other FCNSI effects for further constraints.
As mentioned,
decays of the type $P^0 \rightarrow \ell^+ \ell^-$ are only relevant
if the scalar has a relevant coupling to $\ell^+ \ell^-$.
In theories where this coupling is free,
it could even vanish and no addition constraint arises.
But in some theories this coupling is fixed by other
quantities (such as masses and, possibly, elements of
the leptonic mixing matrix),
and they must be taken into account.
For completeness,
the appendix includes the relevant formulae.

Finally,
if the FCNSI couplings are large,
they could induce
$S \rightarrow q_i \bar{q}_j$ decays.
Using the expressions in the appendix
and the values of $a^\textrm{max}$ and $b^\textrm{max}$ in Table~\ref{TAB:MesonMixing:03}
for $m_S=125 \textrm{GeV}$,
we find that the effects for $S \rightarrow \bar{b} s + \bar{s} b$ are less
than a few percent of the total width,
and completely irrelevant for the other FCNSI decays.
As seen in Fig.~\ref{fig:2},
for a sufficiently large $m_S$ the maximum scalar partial width
of $S \rightarrow \bar{b} s + \bar{s} b$ could exceed
$S \rightarrow \bar{b} b $.
But for masses above $\sim 135\textrm{GeV}$, the decays into vector bosons
take over.
As a result, the $S \rightarrow \bar{b} s + \bar{s} b$ decay
will typically be a minute fraction of the total width.

\section{\label{sec:conclusions}Conclusions}

In general,
a theory with more than one neutral scalar will induce FCNSI.
Because these lead to mixing in the neutral meson systems,
it is customary to eliminate or suppress such contributions.
Strategies in the literature range from discrete symmetries --
for example, a $Z_2$ symmetry in 2HDMs \cite{Glashow:1976nt, Paschos:1976ay} --,
to relation with CKM matrix elements \cite{BGL},
and/or large scalar masses.
In this article,
we highlight a further possibility,
which could occur for a spin 0 state
with both scalar and pseudoscalar couplings.
In that case,
one could have a self-cancelation between both couplings.

We explain in detail how this mechanism could occur even in models
where the Higgs potential and the vacuum preserve CP.
We illustrated this effect by showing the explicit
couplings of the BGL models \cite{BGL}.
This case shows that a self-cancelation is quite natural,
at least to some degree,
depending on the exact values of the hadronic matrix elements.
Our evaluations of this effect in the $B$ systems are more
promising using lattice estimates than using estimates
with the vacuum insertion approximation.

We also investigated the possibility that such FCNSI could
induce a new mechanism of scalar production at LHC.
In a theory with multi Higgs we expect that FCNSI can be ignored
as a production mechanism,
except in two cases: for some scalar where self-cancelation
is active to some accuracy;
or in limiting cases where the contribution from different scalars
cancel each other, as illustrated in
at the end of Section~\ref{subsec:CPcons}.
These effects should not be ruled out a priori.

\vspace{1ex}

\begin{acknowledgments}
J.P.S. is very grateful to Lincoln Wolfenstein for many discussions
on the topics covered in this article,
who sadly passed away recently.
We are grateful to F.J. Botella,
L. Lavoura, P. Pal, J. Rom\~{a}o, and R. Santos for discussions.
This work is supported in part by the Portuguese
\textit{Funda\c{c}\~{a}o para a Ci\^{e}ncia e Tecnologia} (FCT)
under contract UID/FIS/00777/2013.
M.N. is supported in part by FCT through a postdoctoral fellowship under
PTDC/FIS-NUC/0548/2012, and through CERN/FP/123580/2011;
these projects are partially funded
through COMPETE, QREN, POCTI (FEDER) and EU.
\end{acknowledgments}

\appendix
\section{\label{app:rates}Further FCNSI processes}

We are also interested in the decays of $P^0$ into two
a pair of charged leptons $P^0 \rightarrow \ell^+ \ell^-$.
We write the couplings of the scalar into the lepton pair as
\be
-{\cal L}_Y
=
\Sh \left\{
\bar{\ell}\, (a_{\ell} + i b_{\ell} \gamma_5)\, \ell
\right\} + \cdots
\label{L_mu}
\ee
where $a_\ell$ and $b_\ell$ are real.
We find\footnote{The $\overline{P^0} \rightarrow \ell^+ \ell^-$ would seem to
	involve the different factor $|b_{{\bar{i}}j}|$, but, given Eq.~\eqref{hermiticity},
	$|b_{{\bar{i}}j}|=|b_{{\bar{j}}i}|$.}
\be
\Gamma[P^0 \rightarrow \ell^+ \ell^-]
=
\frac{1}{8 \pi}
\frac{m_P^5\, f_P^2}{(m_{q_j} + m_{q_i})^2\ m_\Sh ^4}\,
|b_{{\bar{j}}i}|^2\,
\beta_\ell\,
\left[
|a_\ell|^2\, \beta_\ell^2 + |b_\ell|^2
\right],
\label{Gamma_ll}
\ee
where
\be
\beta_\ell = \sqrt{1 - \frac{4 m_\ell^2}{m_P^2}},
\ee
and the form factor $f_P$ arises from
\be
\langle 0 |
\bar{q}_j \gamma_5 q_i
| P^0 \rangle
= i \frac{m_P^2}{m_{q_j} + m_{q_i}} \, f_P,
\hspace{6ex}
\langle 0 |
\bar{q}_j q_i
| P^0 \rangle = 0.
\label{ME}
\ee
Notice that $a_{{\bar{j}}i}$ does not appear in Eq.~\eqref{Gamma_ll} because
the pseudoscalar meson $P$ cannot couple to the scalar component of $\Sh $,
as seen in the second matrix element of Eq.~\eqref{ME}.

Finally,
the Lagrangian in Eq.~\eqref{L} induces $h \rightarrow \bar{q}_i q_j$
decays.
We find
\be
\Gamma \left[ \Sh  \rightarrow \bar{q}_i q_j\right]
=
\frac{N_c}{8 \pi\, m_\Sh ^3}
\sqrt{\alpha_+ \alpha_-}
\left[
|a_{{\bar{j}}i}|^2 \alpha_+
+
|b_{{\bar{j}}i}|^2 \alpha_-
\right],
\ee
where
\be
\alpha_\pm = m_\Sh ^2 -
\left(
m_{q_i} \pm m_{q_j}
\right)^2.
\ee


\vspace{2ex}


\begin{thebibliography}{99}
%
\bibitem{ATLASHiggs}
G.~Aad {\it et al.}  [ATLAS Collaboration],
  Phys.\ Lett.\ B {\bf 716}, 1 (2012)
  [arXiv:1207.7214 [hep-ex]].
%
\bibitem{CMSHiggs}
S.~Chatrchyan {\it et al.}  [CMS Collaboration],
  Phys.\ Lett.\ B {\bf 716}, 30 (2012)
  [arXiv:1207.7235 [hep-ex]].
%
\bibitem{ALEPH:2005ab}
  S.~Schael {\it et al.}  [ALEPH and DELPHI and L3 and OPAL and SLD and LEP Electroweak Working Group and SLD Electroweak Group and SLD Heavy Flavour Group Collaborations],
  Phys.\ Rept.\  {\bf 427}, 257 (2006)
  [hep-ex/0509008].
%
\bibitem{hhg}
  J.F.~Gunion, H.E.~Haber, G.L.~Kane and S.~Dawson,
  \textit{The Higgs Hunter's Guide}
  \mbox{(Westview Press, Boulder, CO, 2000)}.
%
\bibitem{ourreview}
G.~C.~Branco, P.~M.~Ferreira, L.~Lavoura, M.~N.~Rebelo, M.~Sher, and J.~P.~Silva,
\emph{Theory and phenomenology of two-Higgs-doublet models},
\emph{Phys.\ Rept.\ }  {\bf 516}, 1 (2012)
[arXiv:1106.0034 [hep-ph]].
%
\bibitem{Christenson:1964fg}
  J.~H.~Christenson, J.~W.~Cronin, V.~L.~Fitch and R.~Turlay,
  Phys.\ Rev.\ Lett.\  {\bf 13}, 138 (1964).
%
\bibitem{Glashow:1970gm}
  S.~L.~Glashow, J.~Iliopoulos and L.~Maiani,
  Phys.\ Rev.\ D {\bf 2}, 1285 (1970).
%
\bibitem{Aubert:2001nu}
  B.~Aubert {\it et al.}  [BaBar Collaboration],
  Phys.\ Rev.\ Lett.\  {\bf 87}, 091801 (2001)
  [hep-ex/0107013].
%
\bibitem{Abe:2001xe}
  K.~Abe {\it et al.}  [Belle Collaboration],
  Phys.\ Rev.\ Lett.\  {\bf 87}, 091802 (2001)
  [hep-ex/0107061].
%
\bibitem{CMS:2014xfa}
  V.~Khachatryan {\it et al.}  [CMS and LHCb Collaborations],
  Nature {\bf 522}, 68 (2015)
  [arXiv:1411.4413 [hep-ex]].
%
\bibitem{Fontes:2015mea}
  D.~Fontes, J.~C.~Rom\~{a}o, R.~Santos and J.~P.~Silva,
  JHEP {\bf 1506}, 060 (2015)
  [arXiv:1502.01720 [hep-ph]].
%
\bibitem{BLS}
 G.~C.~Branco, L.~Lavoura and J.~P.~Silva,
 \textit{``CP Violation''},
 Oxford University Press,
 Int.\ Ser.\ Monogr.\ Phys.\  \textbf{103} (1999) 1.
%
\bibitem{Jung:2013hka}
  M.~Jung and A.~Pich,
  JHEP {\bf 1404}, 076 (2014)
  [arXiv:1308.6283 [hep-ph]].
%
\bibitem{ckmfitter}
CKMfitter Group (J.~Charles \textit{et al.}),
Eur.\ Phys.\ J.\ C {\bf 41}, 1 (2005) [hep-ph/0406184].
Updated results and plots available at: http://ckmfitter.in2p3.fr.
%
\bibitem{Lenz:2012az}
  A.~Lenz, U.~Nierste, J.~Charles, S.~Descotes-Genon, H.~Lacker, S.~Monteil, V.~Niess and S.~T'Jampens,
  Phys.\ Rev.\ D {\bf 86}, 033008 (2012)
  [arXiv:1203.0238 [hep-ph]].
%
\bibitem{Bona:2007vi}
  M.~Bona {\it et al.} [UTfit Collaboration],
  JHEP {\bf 0803}, 049 (2008)
  [arXiv:0707.0636 [hep-ph]].
  Updated results and plots available at: http://www.utfit.org/.
%
\bibitem{Agashe:2014kda}
  K.~A.~Olive {\it et al.} [Particle Data Group Collaboration],
  Chin.\ Phys.\ C {\bf 38}, 090001 (2014).
%
\bibitem{Bouchard:2011xj}
C.~M.~Bouchard, E.~D.~Freeland, C.~Bernard, A.~X.~El-Khadra, E.~Gamiz, A.~S.~Kronfeld, J.~Laiho and R.~S.~Van de Water,
PoS LATTICE {\bf 2011}, 274 (2011)
[arXiv:1112.5642 [hep-lat]].
%
\bibitem{LS}
L.\ Lavoura and J.\ P.\ Silva,
Phys.\ Rev.\ D \textbf{50}, 4619 (1994)
  [arXiv:9404276 [hep-ph]].
%
\bibitem{BS}
F.~J.~Botella and J.~P.~Silva,
  Phys.\ Rev.\  D {\bf 51}, 3870 (1995)
  [arXiv:hep-ph/9411288].
%
\bibitem{Ferreira:2010bm}
  P.~M.~Ferreira and J.~P.~Silva,
  Eur.\ Phys.\ J.\ C {\bf 69}, 45 (2010)
  [arXiv:1001.0574 [hep-ph]].
%
\bibitem{care}
Some care is needed when using $h$ and $H$ from different sources,
especially when interfering signs matter.
The different definitions in the literature may be written as
%
\be
\left(
\begin{array}{c}
\xi_H\, H\\
\xi_h\, h
\end{array}
\right)
=
\left(
\begin{array}{cc}
c_\alpha & s_\alpha\\
- s_\alpha & c_\alpha
\end{array}
\right)
\left(
\begin{array}{c}
\rho_1\\
\rho_2
\end{array}
\right)
\ee
%
where $\xi_h$ and $\xi_H$ represent possible independent choices
for the signs $\pm$.
Here we use $\xi_h = \xi_H = +$,
while, for example,
Eqs.~(12)-(14) of Ref~\cite{ourreview} use $\xi_h = \xi_H = -$.
%
\bibitem{BGL}
G.~C.~Branco, W.~Grimus and L.~Lavoura,
Phys.\ Lett.\ B {\bf 380}, 119 (1996)
[hep-ph/9601383].
%
\bibitem{Botella:2014ska}
F.~J.~Botella, G.~C.~Branco, A.~Carmona, M.~Nebot, L.~Pedro and M.~N.~Rebelo,
JHEP {\bf 1407}, 078 (2014)
[arXiv:1401.6147 [hep-ph]].
%
\bibitem{Ferreira:2010ir}
P.~M.~Ferreira and J.~P.~Silva,
Phys.\ Rev.\ D {\bf 83}, 065026 (2011)
[arXiv:1012.2874 [hep-ph]].
%
\bibitem{Botella:2011ne}
  F.~J.~Botella, G.~C.~Branco, M.~Nebot and M.~N.~Rebelo,
  JHEP {\bf 1110}, 037 (2011)
  [arXiv:1102.0520 [hep-ph]].
%
\bibitem{WolfCKM}
  L.~Wolfenstein,
  Phys.\ Rev.\ Lett.\  {\bf 51}, 1945 (1983).
%
\bibitem{Harlander:2012pb}
  R.~V.~Harlander, S.~Liebler and H.~Mantler,
  Comput.\ Phys.\ Commun.\  {\bf 184}, 1605 (2013)
  [arXiv:1212.3249 [hep-ph]].
  Updated files and manuals available at: http://sushi.hepforge.org/.
%
\bibitem{Dicus:1988cx}
  D.~A.~Dicus and S.~Willenbrock,
  Phys.\ Rev.\ D {\bf 39}, 751 (1989).
%
\bibitem{Balazs:1998sb}
  C.~Balazs, H.~J.~He and C.~P.~Yuan,
  Phys.\ Rev.\ D {\bf 60}, 114001 (1999)
  [hep-ph/9812263].
%
\bibitem{Maltoni:2003pn}
  F.~Maltoni, Z.~Sullivan and S.~Willenbrock,
  Phys.\ Rev.\ D {\bf 67}, 093005 (2003)
  [hep-ph/0301033].
%
\bibitem{Harlander:2003ai}
  R.~V.~Harlander and W.~B.~Kilgore,
  Phys.\ Rev.\ D {\bf 68}, 013001 (2003)
  [hep-ph/0304035].
%
\bibitem{Harlander:2011aa}
  R.~Harlander, M.~Kramer and M.~Schumacher,
  arXiv:1112.3478 [hep-ph].
%
\bibitem{Dittmaier:2011ti}
  S.~Dittmaier {\it et al.}  [LHC Higgs Cross Section Working Group
Collaboration],
  arXiv:1101.0593 [hep-ph].
%
\bibitem{Dittmaier:2012vm}
  S.~Dittmaier, S.~Dittmaier, C.~Mariotti, G.~Passarino, R.~Tanaka,
S.~Alekhin, J.~Alwall and E.~A.~Bagnaschi {\it et al.},
  arXiv:1201.3084 [hep-ph].
%
\bibitem{website}
https://twiki.cern.ch/twiki/bin/view/LHCPhysics/CERNYellowReportPageAt8TeV.
%
\bibitem{Glashow:1976nt}
  S.~L.~Glashow and S.~Weinberg,
  Phys.\ Rev.\ D {\bf 15}, 1958 (1977).

\bibitem{Paschos:1976ay}
  E.~A.~Paschos,
  Phys.\ Rev.\ D {\bf 15}, 1966 (1977).
%
%
%
\end{thebibliography}
\end{document}